\documentclass[aps,prb,twocolumn,amssymb,amsfonts,groupedaddress,showpacs]{revtex4}
\usepackage{epsfig}
\usepackage{amsmath}
\usepackage{bm}
\usepackage{times}

\begin{document}
\title{Decoherence of spin qubits due to a nearby charge fluctuator in gate-defined double dots}

\author{Guy Ramon}
\email{gramon@scu.edu}
\affiliation{Department of Physics, Santa Clara University, Santa Clara, CA 95053}
\author{Xuedong Hu}
\affiliation{Department of Physics, University at Buffalo, SUNY, Buffalo, NY
14260-1500}

\begin{abstract}

The effects of a nearby two-level charge fluctuator on a double-dot two-spin qubit are studied theoretically. Assuming no direct tunneling between the charge fluctuator and the qubit quantum dots, the Coulomb couplings between the qubit orbital states and the fluctuator are calculated within the Hund-Mulliken framework to quadrupole-quadrupole order in a multipole expansion.
We identify and quantify the coupling term that entangles the qubit to the fluctuator and analyze qubit decoherence effects that result from the decay of the fluctuator to its reservoir.
Our results show that the charge environment can severely impact the performance of spin qubits, and indicate working points at which this decoherence channel is minimized.
Our analysis also suggests that an ancillary double-dot can provide a convenient point for single-qubit operations and idle position, adding flexibility in the quantum control of the two-spin qubit.

\end{abstract}

\pacs{03.67.Lx, 73.21.La, 85.35.Gv, 85.75.-d}

\maketitle

\section{Introduction}
\label{intro}

The wide-spread interest in quantum information processing in recent years has been a critical driving force in the research of electron spins localized in semiconductor quantum dots (QDs).\cite{HanRMP}
While these two-level systems are attractive candidates for implementation of scalable systems due to their compatibility with conventional microelectronic technology, their quantum control at the single- and
few-qubit level remains a significant experimental challenge. Furthermore, as all solid state systems, they are inherently less isolated from their environment as compared with atomic systems.

An attractive platform to study quantum control and the related problem of decoherence is the system of gate-defined lateral QDs, in which several of the major breakthroughs in spin qubit technology have emerged in recent years. While isolating a single electron in a QD was achieved only in 2000,\cite{Ciorga} rapid progress has been made since then. Long singlet-triplet relaxation times of the order of milliseconds were measured for a single dot,\cite{Hanson} and a lower bound on the spin coherence time (dominated by pure dephasing) exceeding $1\mu s$ was established, using spin-echo techniques in a double dot system.\cite{Petta} The single-spin relaxation and decoherence time scales have since been pushed to the order of 1 s \cite{Amasha} and 0.1 ms\cite{YacobyPC} respectively.

The relative isolation of QD electron spins, which is indicated by these long coherence times, renders their manipulation and readout particularly challenging. This is accomplished by using Pauli spin blockade to convert spin to charge information so that fast measurement of spin states becomes possible.\cite{Ono,Elzerman,DiCarlo} In addition, coherent exchange of two-electron spins in a double dot system,\cite{Petta} and driven Rabi oscillations of single electron spins using oscillating magnetic\cite{Koppens} and electric\cite{Nowack} fields have been demonstrated as well.

Electron spin relaxation via spin-orbit interaction was shown to be an insignificant decoherence channel,\cite{Golovach} and it has been generally accepted that the nuclear spins in the surrounding host material are the main source for the electron spin decoherence in III-V host materials such as GaAs and InAs.\cite{Merkulov} This has led to intensive experimental\cite{Petta,Johnson,Koppens1,Bayer,Tarucha,Steel} and theoretical\cite{Coish,Klauser,Yao,Liu,Witzel,Witzel2,Deng,Lukin} studies of the nuclear environment, and various proposals for alleviating its adverse effects on the electron spin qubit, among which dynamical nuclear spin polarization was suggested\cite{BurLos,HuSSC,Ramon} and demonstrated.\cite{PettaPol,Reilly,Foletti}

In contrast, the effects of the charge environment on QD spin qubits have only recently started to receive some theoretical attention.\cite{Coish,Hu,Culcer} Charge noise in lateral gated devices can originate from various sources. Suggested mechanisms include gate leakage currents via localized states, charge traps near the quantum point contacts (QPCs), donor centers near the gate surface, Johnson noise from the gate electrodes, and switching events in the doping layer, typically located at an interface 100 nm below the surface.\cite{Petta,Pioro} Measurement of the background charge fluctuation in GaAs quantum dots has shown a linear temperature dependence characteristic of $1/f$ noise.\cite{Jung} Random telegraph noise in GaAs lateral gated structures was measured and characterized by Pioro-Ladri\`{e}re {\it et al.}.\cite{Pioro} This noise was attributed to electrons that tunnel from the gate and are trapped near the QPC, causing fluctuations in the conductance with typical frequency of ~ 1 Hz. Applying a positive gate bias during the device cooldown significantly reduces the noise by reducing the density of ionized donors near the surface, thereby suppressing the electron tunneling.\cite{Pioro} Furthermore, background charge fluctuations were suggested as a possible source for the bistable behavior observed in the coupled electron-nuclear spin system,\cite{Koppens1} and telegraph noise induced by the QPCs was also measured recently in double and triple coupled QDs.\cite{Taubert}

 Generally, single-spin qubits in solids rely on the exchange interaction to perform fast two-qubit operations. Furthermore, a number of recent works have utilized two-spin singlet and unpolarized triplet states in biased configuration to encode a logical qubit, which offer better control as compared with single spin states.\cite{Taylor} However, such exchange-coupled spin qubits are vulnerable to dephasing induced by charge noise, since exchange coupling is electrostatic in nature, and singlet and triplet states generally have different charge distributions.\cite{Hu} In the same spirit, the effects of charge noise on the coherence of spin qubits in Silicon double dots were studied very recently.\cite{Culcer} In addition, the effects of a single, randomly positioned, charge impurity on a three-spin encoded qubit in a triple QD were studied by calculating the impurity-induced changes in the qubit orbital levels.\cite{Puerto} Finally, electron-phonon interaction can also lead to dephasing in an exchange coupled double quantum dot because two-spin singlet and triplet states have different charge distributions.\cite{HuArx,Roszak}

There are many types of charge impurities and defects that can generate electrical fluctuations that affect spin qubits in solid states. In this paper we carry out a microscopic calculation focused on the Coulomb coupling between a biased two-spin qubit and a nearby trapped charge fluctuator represented by a two-centered two-level-system (TLS), utilizing a multipole expansion up to and including the quadrupole-quadrupole order. One scenario for such a two-center defect may be for an electron to be trapped around two donor nuclei that have potential wells somewhat lower than other donor nuclei nearby, so that this electron would oscillate between the sites until the charge motion is relaxed by the background charge fluctuations or phonon emissions. Using a master equation formalism we use the calculated qubit-TLS couplings to study the dynamics of the open system that is formed by the spontaneous emission of the TLS coupled to a reservoir. Thus we obtain quantitative estimates of the decoherence and dephasing effects on the spin qubit during various gate operations, and when idle. This analysis enables us to determine optimal working points at which the qubit's sensitivity to charge fluctuations is reduced.

It is important to note that this work is only an initial step in the quantitative analysis of the effects of charge fluctuations on spin qubits. The focus here is on a quantitative evaluation of the qubit-TLS entangling term. We are particularly interested in clarifying how the TLS-qubit coupling could lead to qubit decoherence due to the background charge fluctuations, with the TLS acting as an intermediary between the spin qubit and the charge environment. The TLS coupling to the environment is dealt at a rudimentary level, serving only to demonstrate the applicability of the presented theory in estimating charge-induced spin decoherence. Building on the results given in this paper, the next step should benefit from the extensive work that has been carried out in recent years on charge-environment-induced decoherence in superconducting qubits.\cite{Astafiev,Martinis,Shnirman,Paladino,Bergli,Paladino1,Faoro}

The paper is organized as follows. In section \ref{Couplings}, we derive the Coulomb coupling between the qubit and the TLS, using the Hund-Mulliken approach to calculate the qubit orbital states and a multipole expansion for the Coulomb interaction. In section \ref{dec} we use these results to study the decoherence effects due to charge fluctuations mediated by the qubit-TLS coupling. After deriving the master equations for the system density matrix in section \ref{Master}, we present and analyze in sections \ref{rot}-\ref{3cycle} the resulting dynamics during various single-qubit operations for singlet-triplet qubits. In section \ref{Sweet} we discuss a convenient working point at which the effective exchange energy is zero and quantify the dephasing time. A summary of our results and a brief discussion on possible extensions of this work are given in section \ref{conc}. In Appendix A we calculate the system concurrence, showing the conditions for qubit-TLS entanglement. In appendix B we provide details of the qubit's orbital Hamiltonian. Appendix C lists the full expressions for the qubit-TLS coupling terms, and finally appendix D presents an analytical solution to the master equation for the case of no TLS tunneling.

\section{qubit-TLS Coulomb coupling}
\label{Couplings}
\setcounter{equation}{0}

We consider the Coulomb interaction between a qubit formed from the singlet and
unpolarized triplet spin states of two electrons in a double dot and a nearby two level system (TLS), assuming no qubit-TLS tunnel coupling. To properly describe a biased double quantum dot, we use the Hund-Mulliken model to calculate the qubit orbital states. With Coulomb interaction being spin-independent, the interaction Hamiltonian can generally be written in the form:\cite{delta}
\begin{equation}
{\cal H}_{\rm int}=-\alpha \sigma_z^{\rm T} \otimes I^Q -\beta I^{\rm T} \otimes \sigma_z^Q +
\gamma \sigma_z^{\rm T} \otimes \sigma_z^Q, 
\label{Hint}
\end{equation}
where we have
\begin{eqnarray}
\alpha &=& \frac{1}{4} \left( V^{{\rm TR}}+V^{{\rm SR}}-V^{{\rm TL}}-V^{{\rm SL}} \right) \nonumber\\
\beta &=& \frac{1}{4} \left(  V^{\rm TR}-V^{{\rm SR}}+V^{\rm TL}-V^{\rm SL} \right)\label{abc} \\
\gamma &=& \frac{1}{4} \left( V^{\rm TR}-V^{\rm SR}-V^{\rm TL}+V^{\rm SL} \right) \nonumber
\end{eqnarray}
Here $V^{ij}$ are the Coulomb matrix elements, where the left superscript denotes the qubit state
(Singlet/Triplet) and the right one denotes the TLS state (Left/Right). While the $\alpha$ coupling should not directly affect the qubit spin state, the $\beta$ coupling effectively renormalizes the qubit exchange energy. The $\gamma$ coupling acts to entangle the qubit and the TLS and therefore leads to qubit spin decoherence when the TLS is coupled to a larger reservoir representing the background charge fluctuations. Appendix A formalizes this last statement, showing that the concurrence\cite{HilWoo} of qubit-TLS system under the time evolution of ${\cal H}_{\rm int}$ is nonzero only for a nonzero $\gamma$ coupling.

The TLS's we consider are sufficiently removed from the spin qubit so that there is no exchange coupling between the spins and the single electron in the TLS. With no electrons tunneling between the qubit and the TLS, the two charge distributions are separated in space, and the Coulomb interaction between them can be described systematically using a multipole expansion approach In the following we use this model to evaluate the Coulomb coupling terms $\alpha , \beta , \gamma$.

\subsection{Hund-Mulliken approach for the qubit orbital Hamiltonian}

The sensitivity of an exchange-coupled spin qubit to a remote charge fluctuator comes from the different charge distributions the singlet and triplet states have.\cite{Coish} Thus we first construct the two-electron orbital states by extending the Hund-Mulliken approach\cite{BurLos} to a biased dot configuration.

We start by approximating the orbitals for the two quantum dots by those of two harmonic wells centered at $\pm a \hat{x}$
\begin{equation}
\phi_{\pm a}({\bf r})=\sqrt{\frac{m \omega}{\pi \hbar}} e^{\pm i \frac{ay}{2 l_B^2}} e^{-\frac{m \omega}{2 \hbar} \left[ (x + \Delta x \mp a)^2+y^2 \right]} \chi(z),
\label{orbital}
\end{equation}
where $\omega=b \omega_0$. Here $\omega_0$ is the harmonic oscillator frequency, and the magnetic compression factor, $b$, is given by $b=\sqrt{1+\omega_L^2/\omega_0^2}$ with the Larmor frequency $\omega_L=eB/2mc$. The phase factor in Eq.~(\ref{orbital}) involving the magnetic length $l_B=\sqrt{\hbar c/eB}$ results from a gauge transformation, and $\Delta x=eE/m \omega^2_0$ is the orbital shift due to the electric field. The $z$ direction wavefunction is taken as the ground state (with associated energy $E_{0z}$) of a finite potential well $V_z$ of width $L_z$
\begin{equation}
\chi(z)={\cal N}_z \left\{ \begin{array}{ll} \cos (k_e z) & |z| \leq L_z/2 \\
\frac{k_e}{\sqrt{k_e^2+\kappa^2}} e^{-\kappa \left( |z|-L_z/2 \right)} & |z|>L_z/2
\end{array} \right.
\label{chiz}
\end{equation}
with $k_e=\sqrt{(2m_e/\hbar^2)E_{0z}}$, $\kappa=\sqrt{(2m_e/\hbar^2)(V_z-E_{0z})}$, ${\cal N}_z=(L_z/2+1/\kappa)^{-1/2}$.

To simplify the Hund-Mulliken calculation, the single particle single-dot states are orthonormalized, $\psi_{\pm a}={\cal N}(\phi_{\pm a}-g \phi_{\mp a})$, where $g=(1-\sqrt{1-s^2})/s$, $s=\langle \phi_a | \phi_{-a} \rangle=e^{-d^2(2b-1/b)}$ is the wavefunction overlap ($d=a/a_B$, with $a_B$ the Bohr radius associated with the harmonic QD confinement potential), and ${\cal N}=(1-2sg+g^2)^{-1/2}$. The orthonormalized orbitals are then used to construct 4 two-particle states: the two doubly occupied singlets, $S(2,0)=\psi_{-a} \psi_{-a}$, $S(0,2)=\psi_{a} \psi_{a}$, the separated singlet state, $S(1,1)=(\psi_{-a} \psi_a + \psi_a \psi_{-a})/\sqrt{2}$, and the separated triplet state, $T\equiv T(1,1)=(\psi_{-a} \psi_a - \psi_a \psi_{-a})/\sqrt{2}$. (we neglect the doubly occupied triplet states as their energy is typically much higher for the gate-defined structure we have in mind.\cite{Johnson,Hanson,Koppens}) In the basis of these two-particle states the orbital Hamiltonian is given as
\begin{equation}
H_{\rm orb}= \left(
\begin{array}{cccc}
\epsilon_{20}^S & X & -\sqrt{2} t_H & 0\\
X & \epsilon_{02}^S & -\sqrt{2} t_H & 0\\
-\sqrt{2} t_H & -\sqrt{2} t_H & \epsilon_{11}^S & 0\\
0 &0 & 0 & \epsilon_{11}^T
\end{array} \right),
\label{Horb}
\end{equation}
where the diagonal elements include the Coulomb interactions, and $t_H$ and $X$ are the single- and double-hopping matrix elements, respectively. Calculational details of the orbital Hamiltonian in a biased configuration are given in appendix B. Figure \ref{Fig1n} shows the energy diagram near the $(1,1)$ to $(0,2)$ charge transition, which results from diagonalization of Eq.~(\ref{Horb}), where we also included the polarized triplet states $T_\pm$ splitted by the Zeeman interaction, $H_Z=g \mu_B {\bf B} \cdot \sum_{i=L,R} {\bf S}_i$, with $g=-0.44$ and $\mu_B$ the Bohr magneton. For this figure and throughout this work we have considered $B=100$ mT ($E_Z=2.5 \mu {\rm eV}$), dot confinement $\omega_0=3 {\rm meV}$ ($a_B=19.5$ nm), and half interdot distance (in $a_B$ units) $d=2.8$, corresponding to the experimental parameters in typical gate-defined double dot systems.\cite{Petta,Koppens} The bias shift in the figure is normalized to the Bohr radius:
\begin{equation*}
\widetilde{\Delta x}=\frac{\Delta x}{a_B}=\frac{eEa_B}{\hbar \omega_0},
\end{equation*}
and it is proportional to the interdot bias gate potential.

\begin{figure}[tbp]
\epsfxsize=0.8\columnwidth
\vspace*{1 cm} \centerline{\epsffile{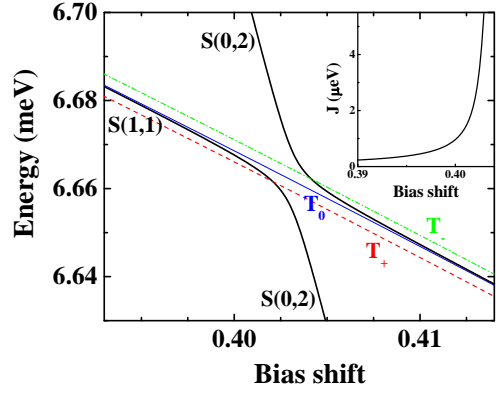}} \vspace{-1.6cm}
\caption{(Color online) Orbital energy diagram for the double dot near the (1,1)-(0,2) transition vs the dimensionless bias shift $\widetilde{\Delta x}$. Shown are the Hybridized singlet states (black curves) and split (1,1) triplet states $T_-$ (dash-dotted green), $T_0$ (solid blue), and $T_+$ (dashed red). The inset shows the exchange energy, calculated as the energy difference between the separated triplet state and the lowest lying hybridized singlet state. Here and throughout the paper $B=100$ mT, $d=2.8$, and $\omega_0=3 {\rm meV}$}
\label{Fig1n}
\end{figure}

Since the orbital Hamiltonian, Eq.~(\ref{Horb}), does not connect the triplet state with any of the singlet states, the combined two-particle orbital-spin triplet state can be written as:
\begin{equation}
|T_0\rangle=\frac{1}{\sqrt{2}} \left( \psi_{-a} \psi_a-\psi_a \psi_{-a} \right) \frac{|\uparrow \downarrow +\downarrow \uparrow \rangle}{\sqrt{2}}.
\end{equation}
The diagonalization of the singlet $3\times 3$ block of $H_{\rm orb}$ yields a hybridized singlet state that is predominantly the separated singlet, $S(1,1)$ at negative or zero bias.\cite{bias} In the basis of the three singlet states $\left\{S(2,0), S(0,2), S(1,1)\right\}$ the lowest-energy orbital-spin singlet state is
\begin{eqnarray}
|S\rangle \!&\!=\!&\! {\cal N}_S \left( \begin{array}{c} a_1\\ a_2 \\ 1 \end{array} \right) \frac{| \uparrow \downarrow - \downarrow \uparrow \rangle}{\sqrt{2}} = {\cal N}_S \left[ a_1 \psi_{-a} \psi_{-a}+a_2 \psi_a \psi_a  \right. \nonumber \\ \!&\!+\!&\! \left. \frac{1}{\sqrt{2}} \left( \psi_{-a} \psi_a +\psi_a \psi_{-a} \right) \right] \frac{|\uparrow \downarrow -\downarrow \uparrow \rangle}{\sqrt{2}}.
\end{eqnarray}
where ${\cal N}_S=1/\sqrt{1+a_1^2+a_2^2}$, and $a_1,a_2$ are the $S(2,0),S(0,2)$ components of the lowest lying singlet eigenstate of the orbital Hamiltonian Eq.~(\ref{Horb}). The exchange energy is defined as the difference between the triplet and this singlet state (see inset of Figure \ref{Fig1n}).

The two-electron states can be expressed in terms of the single particle orthonormal states $\psi_1=(\psi_{-a}+\psi_a)/\sqrt{2}$, $\psi_2=(\psi_{-a}-\psi_a)/\sqrt{2}$, which are more convenient when calculating the qubit-TLS couplings. The two-particle orbital-spin states are given by the slater determinants:
\begin{eqnarray*}
|1_{1\uparrow} 1_{1\downarrow} \rangle \!&\!=\!&\! \frac{1}{\sqrt{2}} \left|
\begin{array}{cc} \psi_1(1) \!\! \uparrow & \psi_1(1) \!\! \downarrow \\
 \psi_1(2) \!\! \uparrow & \psi_1(2) \!\! \downarrow
 \end{array} \right|
 =\psi_1(1)\psi_1(2) \frac{|\uparrow \downarrow -\downarrow \uparrow \rangle}{\sqrt{2}} \\
 |1_{2\uparrow} 1_{2\downarrow} \rangle \!&\!=\!&\! \frac{1}{\sqrt{2}} \left|
\begin{array}{cc} \psi_2(1) \!\! \uparrow & \psi_2(1) \!\! \downarrow \\
 \psi_2(2) \!\! \uparrow & \psi_2(2) \!\! \downarrow
 \end{array} \right|
 =\psi_2(1)\psi_2(2) \frac{|\uparrow \downarrow -\downarrow \uparrow \rangle}{\sqrt{2}}   \\
|1_{1\uparrow} 1_{2\downarrow} \rangle \!&\!=\!&\! \frac{1}{\sqrt{2}} \left|
\begin{array}{cc} \psi_1(1) \!\! \uparrow & \psi_2(1) \!\! \downarrow \\
 \psi_1(2) \!\! \uparrow & \psi_2(2) \!\! \downarrow
 \end{array} \right| \\
\!&\! =\!&\! \frac{1}{\sqrt{2}} \left[\psi_1(1)\psi_2(2)|\uparrow \downarrow \rangle -\psi_2(1)\psi_1(2)|\downarrow  \uparrow \rangle \right],
\end{eqnarray*}
where the number index on the left-hand-side denotes the first or second electron. The triplet and hybridized singlet states can be built as
\begin{eqnarray}
| T_0 \rangle \!&\!=\!&\! \frac{1}{\sqrt{2}} \left( \psi_{-a} \psi_a -\psi_a \psi_{-a} \right) \frac{|\uparrow \downarrow +\downarrow \uparrow \rangle}{\sqrt{2}} \nonumber \\
\!&\!=\!&\! -\frac{1}{\sqrt{2}} \left[ |1_{1\uparrow} 1_{2\downarrow} \rangle+|1_{1\downarrow} 1_{2\uparrow} \rangle \right] \label{T} \\
|S \rangle \!&\!=\!&\! {\cal N}_S \left[ a_1 \psi_{-a} \psi_{-a} +a_2 \psi_a \psi_a +\frac{1}{\sqrt{2}} \left( \psi_{-a} \psi_a+\psi_a \psi_{-a} \right) \right] \nonumber \\
\!&\! \times \!&\! \frac{|\uparrow \downarrow -\downarrow \uparrow \rangle}{\sqrt{2}} \nonumber \\
\!&\!=\!&\! \frac{{\cal N}_S}{\sqrt{2}} \left[ \left(1+\frac{a_1+a_2}{\sqrt{2}}\right) |1_{1\uparrow} 1_{1\downarrow} \rangle - \left(1-\frac{a_1+a_2}{\sqrt{2}}\right) \right. \nonumber \\
\!&\! \times \!&\! \left. |1_{2\uparrow} 1_{2\downarrow} \rangle + \frac{a_1-a_2}{\sqrt{2}}\left( |1_{1\uparrow} 1_{2\downarrow} \rangle-|1_{1\downarrow} 1_{2\uparrow} \rangle \right) \right]. \label{S}
\end{eqnarray}
Notice that the last term in Eq.~(\ref{S}) vanishes for unbiased double dot ($a_1=a_2$). In what follows, these states will be used to calculate the qubit-TLS coupling terms, Eqs.~(\ref{abc}).

\subsection{Multipole expansion for the qubit-TLS interaction}

We model the TLS as a single electron moving in a double well, each of which has a wave function similar to those of the qubit orbitals, Eq.~(\ref{orbital})
\begin{equation}
\phi^{R/L}_T=\frac{1}{\sqrt{\pi} D_T} e^{-\frac{1}{2D_T^2} \left[ \left(x \mp a_T\right)^2 +y^2 \right]} \chi_T (z),
\label{phiT}
\end{equation}
where $D_T$ is the Bohr radius of the (identical) TLS centers, $a_T$ is half the distance between them, and $\chi_T(z)$ is the TLS ground state $z$ wavefunction similar to that of the qubit, Eq.~(\ref{chiz}), with potential $V_z^T$, and width $L_z^T$.\cite{spherical}

The most general Coulomb interaction operator between the qubit and the TLS is given by
\begin{equation}
f_{ijkl}=\int d{\bf r} d{\bf r}' \frac{\rho_Q^{ij}({\bf r}) \rho_{\rm T}^{kl}({\bf r}')}{\varepsilon |{\bf r}-{\bf r}'|}
\end{equation}
where
\begin{eqnarray}
\rho_Q^{ij}({\bf r})&=&e\psi_i^*({\bf r}) \psi_j ({\bf r}) \nonumber \\
\rho_T^{kl}({\bf r})&=&e\phi^{k*}_{T}({\bf r}) \phi^l_{T} ({\bf r}),
\end{eqnarray}
are the electron charge density operators for the qubit ($Q$) and TLS ($T$), where $i,j \in \{1,2\}$ denote the qubit orbital state (symmetric or antisymmetric combination), and $k,l \in \{L,R\}$ denote the TLS state. We use the dielectric constant for GaAs, $\varepsilon =13.1$, and consider only static dielectric constant for screening, since we assume the space near the double dot is completely depleted (i.e. no nearby 2DEG). In addition, we take the TLS inter-site distance to be sufficiently large so as to have a relatively small tunnel coupling, limiting our study to slow TLS's. We can therefore neglect contributions to the qubit-TLS coupling coming from off-diagonal TLS charge densities.\cite{TLStun} To reduce clutter we thus write $f_{ijk} \equiv f_{ijkk}$.

The Coulomb matrix elements of interest are $V^{Tk}= \langle T k|C|Tk \rangle$, and $V^{Sk}=\langle S k|C|S k \rangle$, where $T,S$ are the triplet and singlet states given in Eqs.~(\ref{T})-(\ref{S}). Assuming the creation operators for the electrons in the QDs commute with those in the TLS (i.e., no tunneling between the qubit and the TLS) we find
\begin{eqnarray}
\langle Tk|C|Tk\rangle \!&\!=\!&\! f_{11k}+f_{22k} \nonumber \\
\langle Sk|C|Sk \rangle \!&\!=\!&\! f_{11k}+f_{22k} + {\cal N}_S^2 (a_1+a_2) \label{ST} \\
\!&\!\times \!&\!  \left[ \sqrt{2} \left(f_{11k}-f_{22k} \right)
 + (a_1-a_2) \left(f_{12k}+f_{21k} \right) \right]. \nonumber
\end{eqnarray}

\begin{figure}[tbp]
\epsfxsize=1\columnwidth
\vspace*{1.5cm} \centerline{\epsffile{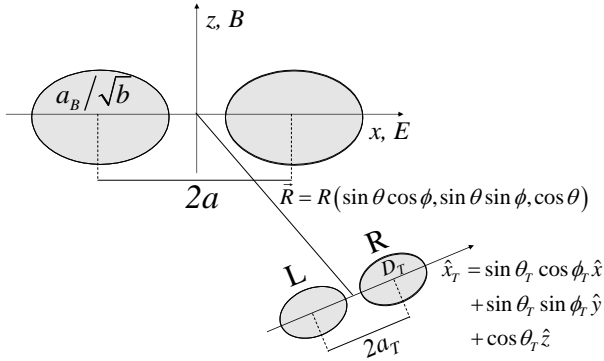}} \vspace{-3.2cm}
\caption{qubit-TLS system geometry.}
\label{Fig2n}
\end{figure}

We calculate the Coulomb interaction terms $f_{ijk}$ by evaluating the electrostatic energy associated with placing the TLS charge distribution in the potential $\Phi_Q^{ij}$, that is due to the qubit charge distribution, expanding the latter in spherical harmonics\cite{Jackson}
\begin{eqnarray}
f_{ijk}\!&\!=\!&\! \int d{\bf r} \Phi_Q^{ij}({\bf r}) \rho_T^{kk}({\bf r}), \nonumber \\
\Phi_Q^{ij}({\bf r})\!&\!=\!&\! \frac{4 \pi}{\varepsilon} \sum_{l=0}^{\infty}\frac{1}{2l+1} \sum_{m=-l}^l\int d{\bf r}' Y_{lm}^*(\theta',\phi')r'^l \rho_Q^{ij}({\bf r}') \nonumber \\
\!&\! \times \!&\! \frac{Y_{lm}(\theta,\phi)}{r^{l+1}} \nonumber \\
\!&\!=\!&\! \frac{q^{ij}_Q}{r}+\frac{{\bf p}^{ij}_Q \cdot {\bf r}}{r^3} +\frac{1}{2} \sum_{lm} Q_{Qlm}^{ij} \frac{r_l r_m}{r^5} + \ldots
\end{eqnarray}
where $\varepsilon$ is the dielectric constant, and $q_Q^{ij},{\bf P}_Q^{ij},Q_Q^{ij}$ are the charge, dipole, and quadrupole electric moments, respectively, associated with the qubit charge distribution.
Combining this with the Taylor expansion for the potential
\begin{equation}
\Phi ({\bf r})=\Phi(0)+{\bf r} \cdot \mbox{\boldmath $\nabla$} \Phi (0)+\frac{1}{2}\sum_{lm} r_l r_m \frac{\partial^2 \Phi}{\partial r_l \partial r_m}(0) + \ldots
\end{equation}
and using $q_T^{k},{\bf P}_T^{k},Q_T^{k}$ to denote the charge, dipole, and quadrupole electric moments, respectively, of the TLS charge distribution, we obtain $f_{ijk}$ up to and including quadrupole-quadrupole order:
\begin{widetext}
\begin{eqnarray}
f_{ijk}\!&\! = \!&\! \Phi_Q^{ij}(R) q^k_T+ \mbox{\boldmath $\nabla$} \left. \Phi^{ij}_Q \right|_{{\bf r}=R} \!\!\!\!\! \cdot {\bf P}^k_T +\frac{1}{6} \sum_{lm} \left. \frac{\partial^2 \Phi^{ij}_Q}{\partial r_l \partial r_m}\right|_{{\bf r}=R} \!\!\!\!\!\! Q^{k}_{Tlm} \notag \\
\!&\!=\!&\! \frac{q_Q^{ij} q_T^k}{\varepsilon R}+\frac{q_T^k {\bf P}_Q^{ij} \cdot \hat{\bf R} +q_Q^{ij} {\bf P}_T^k \cdot \hat{\bf R}}{\varepsilon R^2} + \frac{{\bf P}_Q^{ij} \cdot {\bf P}_T^k -3 \left( {\bf P}_Q^{ij} \cdot \hat{\bf R}\right) \left({\bf P}_T^k \cdot \hat{\bf R} \right)}{\varepsilon R^3} + \frac{1}{2} \sum_{lm}\frac{ q_T^k Q_{Qlm}^{ij} +q_Q^{ij} Q_{Tlm}^k}{\varepsilon R^5} R_l R_m \notag \\
\!&\!+\!&\!\left[ \sum_{lm} \left( P_{Tl}^k Q_{Qlm}^{ij} +P_{Qm}^{ij} Q_{Tlm}^k \right) \frac{R_l}{\varepsilon R^5} - \frac{5}{2} \sum_{lm}\left(Q_{Qlm}^{ij} {\bf P}_T^k \cdot \hat{\bf R}+ Q_{Tlm}^k {\bf P}_Q^{ij} \cdot \hat{\bf R} \right) \frac{R_l R_m}{\varepsilon R^6} \right] \notag \\
\!&\!+\!&\! \left[ \frac{1}{6} \sum_{lm} \frac{ Q_{Qlm}^{ij} Q_{Tlm}^k} {\varepsilon R^5} - \frac{5}{6} \sum_{lmn} \left( Q_{Qnm}^{ij} Q_{Tlm}^k +Q_{Qlm}^{ij} Q_{Tnm}^k  \right) \frac{R_n R_l}{\varepsilon R^7} +\frac{35}{12} \sum_{lmns} Q_{Qns}^{ij} Q_{Tlm}^k \frac{R_l R_m R_n R_s}{\varepsilon R^9} \right].
\label{multipole}
\end{eqnarray}
\end{widetext}
In Eq.~(\ref{multipole}) ${\bf R}= R \left( \sin \theta \cos \phi, \sin \theta \sin \phi, \cos \theta \right)$ is the vector connecting the qubit and TLS centers, and the two dots lie along the $X$-axis. The centers of the TLS are aligned along the axis $\hat{x}_T=\sin \theta_T \cos \phi_T \hat{x}+\sin \theta_T \sin \phi_T \hat{y}+\cos \theta_T \hat{z}$, so the angular dependence of the qubit-TLS interaction is specified by the four angles $(\theta, \phi, \theta_T, \phi_T)$. The system geometry is depicted in Figure \ref{Fig2n}.

The electrical monopole (charge) for both the qubit and TLS is just $e$.\cite{monopole} The qubit dipole moments are, by construction, in the $x$ direction and are found to be
\begin{equation}
{\bf P}_{Q}^{ii}=\int d{\bf r} {\bf r} \rho_Q^{ii}=-\frac{e a_B(1\mp g)^2 \widetilde{\Delta x}}{1-2sg+g^2} (1\pm s) \hat{x},
\end{equation}
where the upper (lower) sign corresponds to $i=1$ ($i=2$). For the mixed qubit orbital Coulomb matrix elements we only need to consider the sum $f_{12k}+f_{21k}$ (see Eq.~(\ref{ST}) for the off-diagonal matrix element).
In the case of the qubit dipole moments this sum is
\begin{equation}
{\bf P}_{Q}^{12}+{\bf P}_{Q}^{21}=-\frac{2ea_B d (1- g^2)}{1-2sg+g^2} \hat{x}.
\end{equation}
Notice that for unbiased dots ($\Delta x=0$) the diagonal qubit dipole moments, $P_{Q}^{11},P_{Q}^{22}$ vanish, since in this case the double dot is symmetric ($\phi_{-a}(-x)=\phi_a(x)$) thus the charge distribution has mirror symmetry around the $y-z$ plane and the dipole moment is identically zero. The introduction of bias allows one of the double occupied states to mix more strongly into the ground singlet state so that a finite dipole moment emerges.

Using the TLS wavefunctions Eq.~(\ref{phiT}) we find the TLS dipole moment as
\begin{equation}
{\bf P}_T^k=\int d{\bf r} {\bf r}_T \rho_T^{kk}=\pm ea_T \hat{x}_T
\end{equation}
where the plus (minus) sign corresponds to $k=R$ ($k=L$).
The quadrupole moments $Q^{ij}_{lm}=\int d{\bf r} (3r_l r_m-r^2 \delta_{lm}) \rho^{ij}({\bf r})$ are found to have only diagonal elements. For the qubit charge distribution they are
\begin{eqnarray}
Q_{Qxx}^{ii}\!&\!=\!&\! \frac{e(1\mp g)^2 a_B^2}{1-2sg+g^2} \left[ (1\pm s)\left(\frac{1}{2b}+2\widetilde{\Delta x}^2-\tilde{l}^2_z+2d^2\right) \right. \nonumber \\
\!&\! \mp \!&\! \left. sd^2 \left(1+\frac{1}{b^2}\right) \right] \notag \\
Q_{Qyy}^{ii}\!&\!=\!&\! \frac{e(1\mp g)^2 a_B^2}{1-2sg+g^2} \left[ (1\pm s)\left(\frac{1}{2b}-\widetilde{\Delta x}^2-\tilde{l}^2_z-d^2\right) \right. \nonumber \\
\!&\! \mp \!&\! \left. sd^2 \left(1-\frac{2}{b^2}\right) \right] \label{Qii} \\
Q_{Qzz}^{ii}\!&\!=\!&\! \frac{e(1\mp g)^2 a_B^2}{1-2sg+g^2} \left[ (1\pm s)\left(-\frac{1}{b}-\widetilde{\Delta x}^2+2\tilde{l}^2_z-d^2\right) \right. \nonumber \\
\!&\! \mp \!&\! \left. sd^2 \left(-2+\frac{1}{b^2}\right) \right] \notag
\end{eqnarray}
and
\begin{eqnarray}
Q_{Qxx}^{12}+Q_{Qxx}^{21} \!&\!=\!&\! 8de \widetilde{\Delta x} a_B^2 \frac{1- g^2}{1-2sg+g^2} \nonumber \\
Q_{Qyy}^{12}+Q_{Qyy}^{21} \!&\!=\!\!&\! Q_{Qzz}^{12}+Q_{Qzz}^{21} \nonumber \\
\!&\!=\!&\! -\frac{1}{2} \left(Q_{Qxx}^{12}+Q_{Qxx}^{21} \right).
\end{eqnarray}
In Eqs.~(\ref{Qii}) the upper (lower) sign refers to $i=1$ ($i=2$), and
\begin{eqnarray}
\tilde{l}_z^2\!&\!=\!&\! \frac{1}{a_B^2}\int dz z^2 \chi^2(z) = \frac{a_B^{-2}}{\kappa L_z+2}\left[\frac{L_z^2}{2}\left(1+\frac{\kappa L_z}{6}\right) \right. \nonumber \\
\!&\!+\!&\! \left. \frac{1}{\kappa^2}(1+\kappa L_z)-\frac{1}{k_e^2} \left(1+\frac{\kappa L_z}{2} \right) \right].
\label{lz2}
\end{eqnarray}
Similarly, the quadrupole moments of the TLS charge distribution are
\begin{eqnarray}
Q_{Txx}^k\!\!&\!\!=\!\!&\!\!  e\left(2a_T^2+\frac{D_T^2}{2}-l_{zT}^2\right) \notag \\
Q_{Tyy}^k\!\!&\!\!=\!\!&\!\! e\left( -a_T^2+\frac{D_T^2}{2}-l_{zT}^2 \right)\\
Q_{Tzz}^k\!\!&\!\!=\!\!&\!\! e\left( -a_T^2-D_T^2 +2l_{zT}^2 \right)\notag
\end{eqnarray}
where $l_{zT}^2=\int dz z^2 \chi_T^2(z)$ is the square extension of the z-direction TLS wavefunction, similar to Eq.~(\ref{lz2}). Note that the TLS quadrupole moment matrix elements are given in the rotated frame $(\hat{x}_T,\hat{y}_T,\hat{z}_T)$ and have the same values for $k=L,R$.

Using the above results we obtain the qubit-TLS coupling terms, Eqs.~(\ref{abc}) to quadrupole-quadrupole order where the nonvanishing contributions are
\begin{eqnarray}
\alpha\!&\!=\!&\!\alpha_{qd}+\alpha_{dd}+\alpha_{Qd} \label{alpha} \\
\beta\!&\!=\!&\!\beta_{dq}+\beta_{Qq}+\beta_{dQ}+\beta_{QQ} \label{beta} \\
\gamma\!&\!=\!&\!\gamma_{dd}+\gamma_{Qd} \label{gamma}.
\end{eqnarray}
The first (second) subscript in each term denotes contribution from the particular multipole moment: monopole (q), dipole (d), and quadrupole (Q) of the qubit (TLS) charge distribution. The explicit expressions for the various coupling terms are rather lengthy and are deferred to appendix C. Note that the dipole-charge, $\beta_{dq}$, dipole-dipole, $\alpha_{dd}$, $\gamma_{dd}$, and dipole-quadrupole, $\beta_{dQ}$ contributions are nonzero only for a biased dot configuration when the qubit dipole moment is nonzero.

\subsection{Qubit-TLS coupling terms}
\label{QTres}

\begin{figure}[tbp]
\epsfxsize=0.7\columnwidth
\vspace*{0.0cm} \centerline{\epsffile{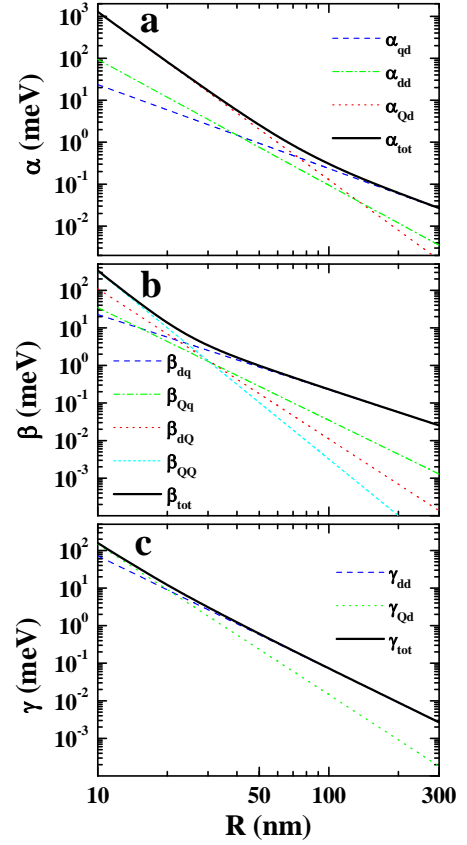}} \vspace{-0.4cm}
\caption{(Color online) Angle-averaged Coulomb coupling terms vs qubit-TLS distance at double-dot bias corresponding to the $(1,1)$ to $(0,2)$ charge transition ($\widetilde{\Delta x}=0.405$).
(a) $\alpha$ terms; (b) $\beta$ terms; (c) $\gamma$ terms.}
\label{Fig3n}
\end{figure}

In order to present graphically the qubit-TLS Coulomb interaction terms, Eqs.~(\ref{alpha})-(\ref{gamma}), we consider a generic system geometry (i.e.~no a-priori knowledge of the relative orientations of the two subsystems). Since giving statistics of various parameters does not shed clear light on the interaction, we give a representative value of the qubit-TLS interaction by averaging over all possible values of qubit-TLS orientation parameters $(\theta, \phi, \theta_T, \phi_T)$. To obtain the correct total values of $\alpha, \beta, \gamma$, this angular averaging should be performed after the addition of the individual contributions from the multipole expansion (i.e., the terms given in Eqs.~(\ref{alpha_qd})-(\ref{gamma_Qd})). Figure \ref{Fig3n} shows the angle-averaged values of the coupling terms vs qubit-TLS distance at the bias point corresponding to the singlet anticrossing (see Figure \ref{Fig1n}). In addition to the double-dot parameters given in Fig.~\ref{Fig1n}, we use here and throughout the paper, unless specified otherwise, QD thickness $L_z=5$ nm, and vertical confinement potential $V_z=500$ meV. For the TLS, we take $L_z^T=3$ nm, $V_z^T=100$ meV, TLS center Bohr radius $D_T=5$ nm, and half TLS centers distance $a_T=20$ nm. The latter are chosen to characterize $\delta$-doped dopants in the insulator with a typical small radius and a fairly large inter-center distance.

The figure demonstrates the convergence of the multipole expansion for each of the three couplings, as $R$ increases. For the above set of parameters we expect higher order contributions in the multipole expansion to be insignificant for qubit-TLS distances exceeding 100 nm (for $\alpha$), 40 nm (for $\beta$), and 30 nm (for $\gamma$). As explained below, it is the $\gamma$ coupling that is responsible for the spin qubit decoherence effects, thus we expect our results to be accurate down to $R=30$ nm.

\begin{figure}[tbp]
\epsfxsize=0.7\columnwidth
\vspace*{-0.0cm} \centerline{\epsffile{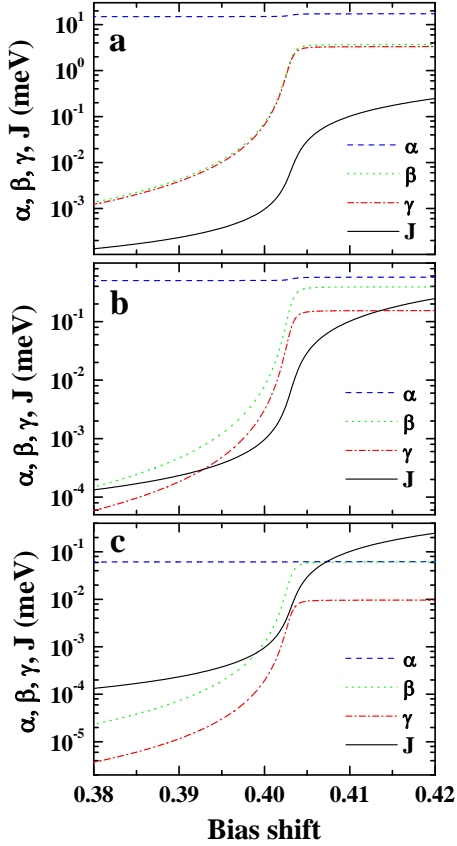}} \vspace{-0.3cm}
\caption{(Color online) Angle-averaged qubit-TLS coupling terms as functions of the double-dot bias for qubit-TLS distances of: (a) $R=30$ nm, (b) $R=80$ nm, and (c) $R=200$ nm. The qubit exchange energy is shown for comparison. All other system parameters are the same as those used in Figure \ref{Fig3n}.}
\label{Fig4n}
\end{figure}

Figure \ref{Fig4n} shows the Coulomb couplings and the qubit exchange energy as functions of the QD bias shift for qubit-TLS distances of $R=30, 80, 200$ nm. While the bias dependence of the $\alpha$ coupling is minimal, both $\beta$ and $\gamma$ strongly depend on the qubit bias, suggesting that the qubit is substantially more susceptible to decoherence due to charge fluctuations at and above the anticrossing point where the $S(0,2)$ component increases significantly in the ground singlet state and the qubit charge distribution acquires a strong dipole component. It is seen that for our parameter choice, $\gamma$ becomes comparable to the exchange energy at $R \lesssim 100 $ nm, below which we anticipate sizable spin decoherence effects due to charge coupling. Inspection of the leading terms in the $\beta$ and $\gamma$ couplings (Eqs.~(\ref{beta_dq})-(\ref{gamma_Qd})) shows that only the latter scales with the TLS centers distance $a_T$, thus the ratio $\beta/\gamma$ decreases with the charge fluctuator size, leading to increased qubit-TLS distances at which the qubit is susceptible to decoherence. We note that the qubit-TLS distance at which the Coulomb terms become appreciable roughly scales linearly with the size of the dots, which is consistent with the basic characteristics of a multipole expansion.

The angular averaging procedure that was used to produce figures \ref{Fig3n} and \ref{Fig4n} was tested by randomly taking values for $(\theta, \phi, \theta_T, \phi_T)$ and using these random 4-vectors to calculate the interaction terms. This calculational mode is useful in later evaluation of decoherence effects such as gate errors. We then calculate the error (or any other decoherence effect) for many randomly selected qubit-TLS geometries and average at the end of the calculation. The results shown in Figs.~\ref{Fig3n} and \ref{Fig4n} were reproduced to within an error $<1\%$ by averaging over 10,000 random runs.

\section{Effects of charge fluctuations on a double-dot spin qubit}
\label{dec}

In this section we examine the effects of the qubit-TLS coupling on the performance of a double-dot two-spin qubit. To do so we employ a master equation formalism to study the dynamics of the coupled qubit-TLS system due to the spontaneous emission of the TLS. Specifically we study the effects of the charge coupling on dephasing of the spin qubit and on the fidelity of specific single-qubit operations.

\subsection{Master equation for the qubit-TLS system}
\label{Master}
We consider the master equation describing the qubit-TLS system, with the TLS coupled to a reservoir that results in its spontaneous emission
\begin{equation}
\frac{d\rho}{dt}=-\frac{i}{\hbar} \left[ {\cal H},\rho \right]+\sum_j \left[2L_j \rho L_j^\dagger-\{L_j^\dagger L_j,\rho\} \right].
\label{lind}
\end{equation}
Here $\rho=\rho_T \otimes \rho_Q$ is the $4 \times 4$ density matrix of the qubit-TLS system, $L_j$ are the Lindblad operators, and the qubit-TLS Hamiltonian is
\begin{equation}
{\cal H}=I^T \otimes {\bf B}_Q \cdot \mbox{\boldmath $\sigma$}^Q +{\bf B}_T \cdot \mbox{\boldmath $\sigma$}^T \otimes I^Q+{\cal H}_{\rm int},
\label{H}
\end{equation}
where ${\cal H}_{\rm int}$ is given in Eq.~(\ref{Hint}), and $Q$ ($T$) superscript denotes an operator on the qubit (TLS) subsystem. In Eq.~(\ref{H}) ${\bf B}_Q=\frac{1}{2} (\delta h,0,J)$, with $\delta h$ the magnetic field inhomogeneity between the dots arising from either application of an inhomogeneous $B$, different $g$ factors in the two dots, or inhomogeneous nuclear polarizations. ${\bf B}_T=(t_T,0,\omega_T)$ where $\omega_T$ is the TLS level splitting and $t_T$ is the tunnel coupling between the two centers. The latter is a function of the TLS Bohr radius $D_T$ and center separation $a_T$, found using Eq.~(\ref{phiT}) to be
\begin{equation*}
t_T= \frac{\hbar^2}{m_e D_T^2} \left(\frac{5}{4}+\frac{a_T^2}{D_T^2} \right) e^{-(a_T/D_T)^2}.
\end{equation*}
We assume coupling of the TLS to a cold bath in the vacuum state through spontaneous emission, described by a single Lindblad operator $L=\sqrt{\Gamma} \sigma_-^T \otimes I^Q$, where $\Gamma$ is the spontaneous emission rate. Measurements of the relaxation time of charge qubits in lateral GaAs double dots yielded $T_1=16$ ns,\cite{PettaPRL} corresponding to $\Gamma \approx 0.04 \mu{\rm eV}^{-1}$.

Transforming Eq.~(\ref{lind}) to the interaction picture, with $\tilde{\rho}(t)=e^{i{\cal H}t} \rho(t) e^{-i{\cal H}t}$, we find
\begin{equation}
\frac{d \tilde{\rho}}{dt}=\Gamma \left[ 2(\tilde{\sigma}_- \tilde{\rho} \tilde{\sigma}_+ -\tilde{\sigma}_+ \tilde{\sigma}_- \tilde{\rho} - \tilde{\rho} \tilde{\sigma}_+ \tilde{\sigma}_- \right]
\label{rhot}
\end{equation}
with
\begin{eqnarray}
\tilde{\sigma}_\pm \!&\!=\!&\! \frac{a_1^{\pm 2}+a_2^{\pm 2}}{2} (\sigma_\pm^T \otimes I^Q)+ \frac{a_1^{\pm 2}-a_2^{\pm 2}}{2} (\sigma_\pm^T \otimes \sigma_z^Q) \nonumber \\
\!&\!\mp \!&\! i \frac{c_1 a_1^\pm +c_2 a_2^\pm}{2} (\sigma_z^T \otimes I^Q) \mp i \frac{c_1 a_1^\pm -c_2 a_2^\pm}{2} (\sigma_z^T \otimes \sigma_z^Q) \nonumber \\
\!&\!+\!&\! \frac{c_1^2+c_2^2}{2} (\sigma_\mp^T \otimes I^Q) +\frac{c_1^2-c_2^2}{2} (\sigma_\mp^T \otimes \sigma_z^Q).
\label{sigma}
\end{eqnarray}
Here we have defined $a_i^\pm=\cos \Omega_i t \pm i(\omega_i/\Omega_i) \sin \Omega_i t$, and $c_i=(t_T/\Omega_i) \sin \Omega_i t$, where $\Omega_i=\sqrt{\omega_i^2+t_T^2}$, $\omega_i=\omega_T-\alpha \pm\gamma$ and the upper (lower) sign corresponds to $i=1$ ($i=2$).

Inserting $\tilde{\sigma}_\pm$ into Eq.~(\ref{rhot}) we find three separable sets of differential equations for the matrix elements of $\tilde{\rho}$. The first set is
\begin{eqnarray}
\frac{\dot{\tilde{\rho}}_{00}}{\Gamma}\!&\!=\!&\!-2 |a_1^+|^4 \tilde{\rho}_{00} +2 c_1^4 \tilde{\rho}_{22}+2c_1 (1-2c_1^2) {\rm Im} (a_1^- \tilde{\rho}_{02}) \nonumber \\
\frac{\dot{\tilde{\rho}}_{22}}{\Gamma}\!&\!=\!&\! 2|a_1^+|^4 \tilde{\rho}_{00} - 2c_1^4 \tilde{\rho}_{22}- 2c_1 (1-2c_1^2) {\rm Im} (a_1^- \tilde{\rho}_{02}) \label{r00} \\
\frac{\dot{\tilde{\rho}}_{02}}{\Gamma}\!&\!=\!&\!-(1+2c_1^2-2c_1^4)\tilde{\rho}_{02}+2c_1^2 a_1^{+2}\tilde{\rho}_{02}^* \nonumber \\
\!&\!+\!&\! ia_1^+c_1 \left[(3-2c_1^2)\tilde{\rho}_{00}+(2c_1^2+1)\tilde{\rho}_{22} \right].\nonumber
\end{eqnarray}
The equations for $\{\tilde{\rho}_{11},\tilde{\rho}_{33},\tilde{\rho}_{13}\}$ take the same form as those above, with $a_1 \rightarrow a_2$, $c_1 \rightarrow c_2$. The third set of equations reads:
\begin{widetext}
\begin{eqnarray}
&\!\!&\frac{1}{\Gamma} \dot{\tilde{\rho}}_{01}=-(2-c_1^2-c_2^2-2c_1 c_2 a_1^- a_2^+) \tilde{\rho}_{01} -i c_2\left(a_2^- -2c_1c_2a_1^-\right) \tilde{\rho}_{03} + ic_1 (a_1^+ -2c_1c_2a_2^+) \tilde{\rho}_{12}^* +2c_1^2 c_2^2 \tilde{\rho}_{23} \nonumber \\
&\!\!& \frac{1}{\Gamma}\dot{\tilde{\rho}}_{03}= ia_2^+(c_2+2c_1a_1^- a_2^+ ) \tilde{\rho}_{01}-\left(1-c_1^2+c_2^2 +2c_1 c_2 a_1^- a_2^+ \right) \tilde{\rho}_{03} + 2c_1^2 a_2^{+2} \tilde{\rho}_{12}^* + ic_1 (a_1^+ +2c_1 c_2a_2^+) \tilde{\rho}_{23} \nonumber \\
&\!\!&\frac{1}{\Gamma}\dot{\tilde{\rho}}_{12}= ia_1^+ (c_1+2c_2 a_1^+ a_2^-) \tilde{\rho}_{01}^*+2c_2^2a_1^{+2} \tilde{\rho}_{03}^* -(1+c_1^2-c_2^2+2c_1 c_2 a_1^+ a_2^-) \tilde{\rho}_{12}+ic_2 (a_2^+ +2c_1 c_2 a_1^+) \tilde{\rho}_{23}^*  \label{r01} \\
&\!\!&\frac{1}{\Gamma}\dot{\tilde{\rho}}_{23}= 2  a_1^{-2} a_2^{+2}\tilde{\rho}_{01}+ia_1^- (2c_2 a_1^-a_2^+ -c_1) \tilde{\rho}_{03} - i a_2^+ (2c_1 a_1^- a_2^+ -c_2) \tilde{\rho}_{12}^* -(c_1^2+c_2^2-2c_1 c_2 a_1^- a_2^+ ) \tilde{\rho}_{23}. \nonumber
\end{eqnarray}
\end{widetext}
These differential equations are analytically solvable only for the case of zero TLS tunneling ($t_T=0$), where closed expressions are also obtained for the matrix elements of the original $\rho(t)$. We include these results in appendix D, as they shed light on several features of the general case. For nonzero TLS tunneling, equations (\ref{r00})-(\ref{r01}) are solved numerically and the resulting interaction-picture density matrix is then transformed back to obtain $\rho(t)$. In what follows we study the effects of a nearby charge fluctuator on a double-dot two electron spin qubit by applying equations (\ref{r00})-(\ref{r01}) to various cases of single qubit rotations.

\subsection{Single qubit rotations}
\label{rot}

In the absence of charge coupling, the time evolution of the qubit state is governed by the first term in the Hamiltonian, Eq.~(\ref{H}), and we obtain a rotation about an axis lying in the $X-Z$ plane whose angle with respect to the $X$ axis is $\vartheta=\arctan(J/\delta h)$. When qubit-TLS coupling is introduced its deteriorating effects largely divide into the effects of the $\beta$ and $\gamma$ interaction terms. The $\beta$ coupling renormalizes the exchange energy, $J \rightarrow \tilde{J}=J-2\beta$, so that we effectively get a rotation about a new axis, whose tilt angle with respect to the $X$ axis is $\tilde{\vartheta}=\arctan[(J-2\beta)/\delta h]$.

\begin{figure}[tbp]
\epsfxsize=0.95\columnwidth
\vspace*{0.0cm} \centerline{\epsffile{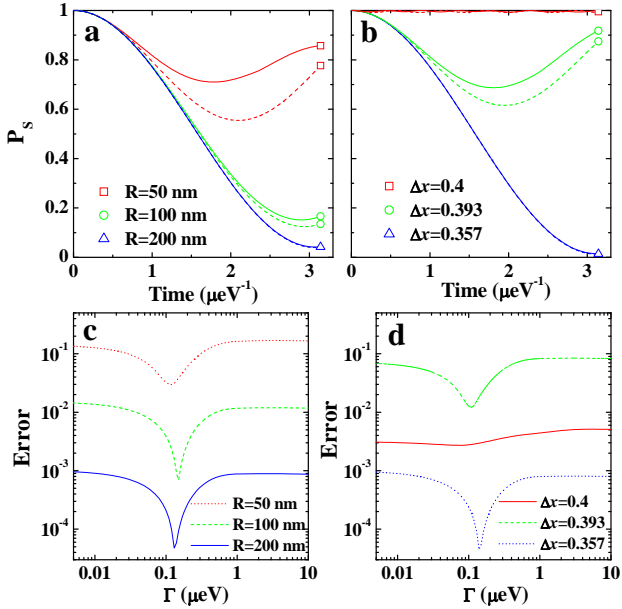}} \vspace{-0.3cm}
\caption{(Color online) (a) Singlet probability vs time for several qubit-TLS distances, at bias $\widetilde{\Delta x}=0.382$. (b) Singlet probability vs time for several bias points at $R=100$ nm. (c) End-of-pulse error relative to the $\beta$ corrected rotation about $\tilde{\vartheta}$-tilted axis vs TLS spontaneous emission rate $\Gamma$, for several $R$ values, at bias $\widetilde{\Delta x}=0.382$. (d) End-of-pulse error vs $\Gamma$ for several bias points at $R=100$ nm. In figures (a) and (b) $\Gamma=0.04 \mu$eV, the solid lines represent the actual time evolution, and dashed lines correspond to pure rotations around the $\tilde{\vartheta}$-tilted axes (see main text). In all the figures pulse duration is taken as $T=\pi/\delta h$ with magnetic field inhomogeneity $\delta h=1 \mu$eV.}
\label{Fig5n}
\end{figure}

Figure \ref{Fig5n} shows the effects of qubit-TLS coupling on several single qubit rotations. In this figure, as well as for results presented in the rest of this paper unless otherwise noted, the coupling terms were calculated by averaging over the qubit-TLS relative orientations $(\theta, \phi, \theta_T, \phi_T)$ as explained in section \ref{QTres}. The system's initial state is a singlet for the qubit and a localized state $|\phi^T_L \rangle$ for the TLS. Other parameters used are: magnetic field inhomogeneity $\delta h=1 \mu$eV, TLS level splitting $\omega_T =5t_T$, and TLS Bohr radius and centers half separation $D_T=5$ nm, $a_T=19.05$ nm, respectively. The resulting TLS tunnel coupling, $t_T \approx 0.36 \mu$eV, is relatively small thus the results can be analyzed in the context of the zero-tunneling analytical solution presented in appendix D.

Figure \ref{Fig5n}a shows the singlet probability as a function of pulse time for several qubit-TLS distances, $R=50, 100, 200$ nm, at a negative bias point $\widetilde{\Delta x}=0.382$ (below the $(1,1)-(0,2)$ transition point), where $J=0.15 \mu$eV. In order to remove the effects of the $\beta$ coupling, which are easily corrected, we compare the resulting time evolution for each $R$ with a rotation about the new tilt-angle $\tilde{\vartheta}$. The dashed lines correspond to $P_S(t)=1-\cos^2\tilde{\vartheta} \sin^2\tilde{B} t$ (see appendix D), and their deviations from the solid curves correspond to the remaining gate errors due to the combined effects of the $\gamma$ coupling and $\Gamma$. The rotation axis is determined by the $\beta$ coupling, thus it varies with $R$. For $R=50, 100, 200$ nm we find $\beta=0.5,0.12,0.03 \mu$eV, respectively, and the corresponding tilt angles are $\tilde{\vartheta}=48^\circ,21^\circ,11^\circ$. Figure \ref{Fig5n}b shows the singlet probabilities vs pulse time for $R=100$ nm at several bias points. Both the exchange and the $\beta$ coupling increase dramatically as the bias turns positive (see Figure \ref{Fig4n}) thus the corresponding tilt angles for $\widetilde{\Delta x}=0.357, 0.393, 0.4$ are $\tilde{\vartheta}=6.5^\circ,52^\circ,85^\circ$ respectively, where the latter is close to rotation about the $Z$ axis (red-lines and square in figure \ref{Fig5n}b).

Figures \ref{Fig5n}c and \ref{Fig5n}d show the pulse error dependence on $\Gamma$, corresponding to Figs.~\ref{Fig5n}a and \ref{Fig5n}b, respectively. For a self-consistent calculation we assume a cubic dependence of $\Gamma$ with the TLS energy splitting, $\Gamma \sim \left( \omega_T^2+(2t_T)^2\right)^{3/2}$, appropriate for free-space spontaneous emission, with the prefactor fixed using the data of ref.~\onlinecite{PettaPRL}. Thus in figures \ref{Fig5n}c and \ref{Fig5n}d, $\Gamma$ variation is accompanied by varying the TLS centers separation resulting in a variation in the TLS tunnel-coupling. We note that due to the exponential dependence of $t_T$ on $a_T$ large variations in $\Gamma$ amount to a very modest change in $a_T$ (8\% change in $a_T$ correspond to three orders-of-magnitude change in $\Gamma$).

Overall we find that the $\gamma$ coupling plays the most important role in determining the pulse errors, while the $\Gamma$ coupling to the reservoir is less significant. The pronounced dips in the pulse error near $\Gamma \sim 0.1 \mu$eV, observed in Figs.~\ref{Fig5n}c,d occur due to accidental matching between the actual and no-$\gamma$ rotations at the end of the pulse, thus they do not reflect the substantial deviations of these rotations that appear throughout the pulse. Naively, one would think that for a very small $\Gamma$, the TLS is coherent and there is no information loss so that gate errors are minimized. While this $\Gamma$ dependence is evident in Figures \ref{Fig5n}c and \ref{Fig5n}d, its effect on our results is marginal as compared with the gate-errors dependence on $\gamma$ (notice that the small reduction of $a_T$ as $\Gamma$ is increased induces a small reduction in $\gamma$ thus the $\Gamma$ dependence in these figures is further masked). Some insight to the secondary role played by $\Gamma$ can be gained from the analytical solution for the $t_T=0$ case given in appendix D.

Inspecting Eq.~(\ref{P_Sanal}) we see that the leading term in the expansion of the pulse error in orders of $\gamma/\tilde{B}$ vanishes for $\tilde{\vartheta}=0^\circ,90^\circ$, and the subleading term that is still present for $\tilde{\vartheta}=0^\circ$, is independent of $\Gamma$. Thus the largest pulse errors are found at $\widetilde{\Delta x}=0.382,R=50$ nm ($14\%-17\%$ error, red squares and solid lines in Figs.~\ref{Fig5n}a,c) and at $\widetilde{\Delta x}=0.393,R=100$ nm ($7\%-8\%$ error, green circles and dashed lines in Figs.~\ref{Fig5n}b,d), where $\tilde{\vartheta}$ is closest to $45^\circ$ and the $\Gamma$ dependence is most pronounced. At negative bias both $\tilde{\vartheta}$ is close to $0^\circ$ (corresponding to rotation about the $X$ axis) and $\gamma$ is very small, leading to a small error (blue dotted line in Fig.~\ref{Fig5n}d). At positive bias $\tilde{\vartheta}$ is fairly close to $90^\circ$, but at the same time $\gamma$ is considerably larger, thus the subleading term contributes appreciably and the dependence in $\Gamma$ is less pronounced (red solid line in Fig.~\ref{Fig5n}d). For yet higher bias, when $\tilde{\vartheta}\rightarrow 90^\circ$ (corresponding to rotation about the $Z$ axis) the smallness of the $\cos^2\tilde{\vartheta}$ in Eq.~(\ref{P_Sanal}) overtakes the increasing value of $\gamma$ and the errors become extremely small.

We stress that our results were obtained  using a simple model for the coupling of the nearby fluctuator with the vacuum, given in terms of amplitude damping. Further investigations are required to determine spin qubit dephasing and gate errors when other forms of coupling of the fluctuator to the charge environment are considered, including specific charge noise spectra.

Notice that single-qubit gates for a two-electron singlet-triplet qubit correspond to two-qubit gates for single-electron single-spin qubits.  Therefore, the results obtained above can be directly applied to single-spin qubits as indications of two-qubit gate errors.  For example, a SWAP gate for single spin qubits is done by turning on the exchange splitting $J$ for a period of time so that $\int J/\hbar dt = \pi$.  This corresponds to a $z$-rotation for a singlet-triplet qubit when $|S+T_0\rangle$ becomes $|S-T_0\rangle$. An error in such a SWAP operation would leave the two single-spin qubits with unwanted entanglement.\cite{Hu}

\subsection{A three-pulse $\pi$ rotation about the $X$ axis}
\label{3cycle}

Next we analyze the effects of charge fluctuations on a scheme proposed by Hanson and Burkard to produce an effective $\pi$-flip about the $X$ axis in the presence of both exchange energy and a fixed inhomogeneous magnetic field $\delta h$.\cite{HanBur} This setting is desirable since it enables us to perform single-qubit operations without relying on a fast control of the interdot tunnel coupling necessary to bring $J$ to zero, an experimentally challenging feat. An arbitrary single-qubit rotation can be obtained with finite exchange energy by applying three successive rotations in the $X-Z$ plane. The three-pulse bias cycle consists of two working points: (i) a negative bias where $J \approx \delta h$ is small (but need not be zero) and the rotation is about a $\vartheta$-tilted axis. (ii) a positive bias in which $J \gg \delta h$ and the rotation is essentially about the $Z$ axis. When $0 \leq \vartheta \leq \pi/4$, an $X$ rotation at an arbitrary angle $\xi$ can be generated by the cycle:
\begin{equation}
U_x (\xi)=U_\vartheta (\chi) U_z (\varphi) U_\vartheta (\chi),
\label{tzt}
\end{equation}
where $U_\vartheta (\chi)$ and $U_z (\varphi)$ are rotation matrices about the $\vartheta$-tilted  and $z$ axes, respectively, and the angles $\chi$ and $\varphi$ are functions of $\vartheta$ and $\xi$.\cite{HanBur}
In particular, a $\pi$-flip about the $X$ axis ($\xi=\pi$) is obtained when:
\begin{equation}
\chi=\arccos (-\tan^2 \vartheta); \hspace{0.5 cm} \varphi=-2\arctan \frac{\sin \vartheta}{\sqrt{\cos 2\vartheta}}.
\label{chiphi}
\end{equation}

Without charge coupling, the pulse durations are fixed so that $t_\vartheta=\chi/\sqrt{\delta h^2+J_\vartheta^2}$, and $t_z=\varphi/\sqrt{\delta h^2+J_z^2}\approx  \varphi/J_z$, where $J_\vartheta$, and $J_z$ are the exchange energies at the two bias points. Typically most of the duty cycle is spent at the negative bias point (i.e., $t_\vartheta \gg t_z$).  By redesigning the pulse sequence to include the renormalized exchange energy , the effects of the $\beta$ coupling can be eliminated. The $\beta$ corrected pulse sequence is obtained by plugging the new tilt-angle $\tilde{\vartheta}=\arctan[(J_\vartheta-2\beta_\vartheta)/\delta h]$ in Eqs.~(\ref{chiphi}), where the $\vartheta$ subscript denote the quantity is evaluated in the first (negative) bias point. For the parameters we are using, depending on the qubit-TLS distance $R$, the angle-averaged ratio $\beta/\gamma$ can be rather small ($\beta/\gamma=2.7, 6.6, 16.6$ for $R=80, 200, 500$ nm, respectively), thus the exchange-renormalization effect of $\gamma$ is non-negligible. We have found that for short qubit-TLS distances ($R \lesssim 160$ nm, $\beta/\gamma<5$), better results are obtained by including the $\gamma$ coupling in the pulse correction, thus this so-called $\beta'$ correction is given by using a tilt-angle $\tilde{\vartheta}'=\arctan[(J_\vartheta-2\beta_\vartheta-2\gamma_\vartheta)/\delta h]$.

\begin{figure}[tbp]
\epsfxsize=0.95\columnwidth
\vspace*{0.0cm} \centerline{\epsffile{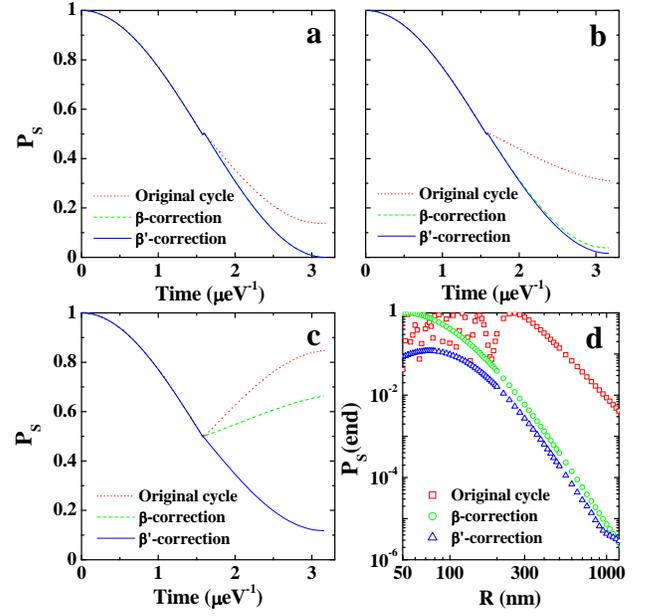}} \vspace{-0.2cm}
\caption{(Color online) Singlet probability as a function of time for original, $\beta$-corrected, and $\beta'$-corrected pulse cycles, with qubit-TLS distance: (a) $R=500$ nm, (b) $R=200$ nm, and (c) $R=80$ nm. (d) $\pi$ flip error as a function of qubit-TLS distance. In all figures $\delta h=1 \mu$eV, $\Gamma=0.04 \mu$eV, $D_T=5$ nm, and $a_T=19.05$ nm. The two bias points in the cycle are: $\widetilde{\Delta x}_\vartheta=0.357$, and $\widetilde{\Delta x}_z=0.414$. The kinks evident in these plots are the result of the three pulses employed in these gate operations.}
\label{Fig6n}
\end{figure}

The above bias cycles were simulated by discretizing bias and time steps, so that the actual switching time between the working points (which should be nonadiabatic) is taken into account.\cite{Switch} We consider a $\pi$-flip rotation with the qubit and TLS initial states taken, as before, to be a singlet and a localized state $|\phi^T_L \rangle$, respectively. Figures \ref{Fig6n}a-c show the singlet probability as a function of cycle time for several qubit-TLS distances, $R=500, 200, 80$ nm, presenting original (dotted-red lines), $\beta$-corrected cycles (dashed-green lines) and $\beta'$-corrected cycles (solid-blue lines). The remaining effects of the $\gamma$ coupling are small (less than 0.1\%) for $R=500$ nm (Fig.~\ref{Fig6n}a), where $\gamma_\vartheta = 0.06$ neV, but grow rapidly with decreasing qubit-TLS distance. As expected, the two corrections deviate only at $R=80$ nm (Fig.~\ref{Fig6n}c).

We notice that the three-pulse cycle may fail for one of two reasons, depending on the $\beta$ coupling. At large $R$ we have $|\beta_z| \approx J_z$ at the positive bias working point. Then for half of the possible system geometries the condition $J_z-2 \beta_z \gg \delta h$ does not hold and we do not obtain the $Z$ rotation necessary to complete the $\pi$ flip. For these geometries, however, there exists a "sweet spot" where $2\beta =J$, thus an effective zero exchange can be obtained at this bias, resulting in a single pulse $X$ rotation. We shall discuss this case in the next section. At small $R$ ($\lesssim 30$ nm for our chosen parameters with $\delta h=1 \mu$eV) a different problem arises due to the large $\beta$ coupling. Since $\beta_\vartheta \gg J_\vartheta$ the corrected tilt angle $\tilde{\vartheta}$ is larger than $\pi/4$ for reasonable values of $\delta h$ and in fact approaches $\pi/2$, as $R$ decreases. An arbitrary-angle $X$ rotation can only be produced when the angle between the axes corresponding to the two working points is between $\pi/4$ and $3\pi/4$, thus, when $\beta$ becomes large, the 3-pulse cycle can generate an $X$ rotation by a maximum angle of
\begin{equation}
\xi_{\rm max}= \left\{ \begin{array}{ll} \arcsin (\cot \tilde{\vartheta)} & \frac{\pi}{4} < \tilde{\vartheta} < \frac{3\pi}{4}\\ \arcsin(-\cot \tilde{\vartheta}) & \frac{5\pi}{4} < \tilde{\vartheta} < \frac{7\pi}{4} \end{array}. \right.
\end{equation}
In order to generate a $\pi$ flip we need to repeat the three pulse cycle, Eq.~(\ref{tzt}), $N_{\rm cyc}={\rm Floor} \left(\frac{\pi}{\xi_{\rm max}} \right)$ times while replacing the rotation angles given in Eq.~(\ref{chiphi}) with $ \chi_{\rm max}=\arccos ( -\cot^2 \tilde{\vartheta})$, $\varphi_{\rm max}=\pi$. Here, $\rm{Floor}(x)$ rounds $x$ to the nearest integer towards minus infinity.
To complete the $\pi$-flip, a final cycle is necessary with a rotation angle $\xi_{\rm f}=\pi-N_{\rm cyc} \xi_{\rm max}$, which is generated with\cite{HanBur}
\begin{eqnarray}
\!\! \chi_{\rm f}\!&\!=\!&\! \arccos \!  \frac{\cos \frac{\xi_{\rm f}}{2} \sqrt{1-\tan^2 \tilde{\vartheta} \sin^2 \frac{\xi_{\rm f}}{2}}-\sin^2 \tilde\vartheta \sin^2 \frac{\xi_{\rm f}}{2}}{\cos^2 \frac{\xi_{\rm f}}{2} +\cos^2 \tilde{\vartheta} \sin^2 \frac{\xi_{\rm f}}{2}} \nonumber \\
\!\! \varphi_{\rm f}\!&\!=\!&\! 2 \pi-2 \arctan \frac{\sin \chi_{\rm f} \sin \tilde{\vartheta}}{\cos^2 \frac{\chi_{\rm f}}{2}+\cos 2 \tilde{\vartheta} \sin^2\frac{ \xi_{\rm f}}{2}}.
\label{chiphif}
\end{eqnarray}

Clearly, as $R$ decreases the $\beta$ coupling becomes larger and more cycles are needed to complete the $\pi$-flip making the process inefficient and more susceptible to other decoherence mechanisms. Indeed, in this regime, the original premise of the bias sequence utilizing two working points with $J_\vartheta \approx \delta h$ and $J_z \ll \delta h$ breaks down and one should consider a different scheme for performing single qubit rotations. As the magnetic field inhomogeneity $\delta h$, used for the qubit rotation decreases, this problem will be manifested at larger qubit-TLS distances. As far as we are concerned, we push down to the small $R$ regime only to demonstrate the increasing effects of the qubit-TLS coupling, although it should be noted that our results are to be taken with caution when $R \lesssim 40$ nm due to slow convergence of the multipole expansion in this regime.

Figure \ref{Fig6n}d shows the singlet probability at the end of the pulse sequence, representing the gate error as a function of qubit-TLS distance for original $\beta-$, and $\beta'-$corrected cycles. As $\gamma$ decreases with increasing $R$, the difference between the two cycles error (green circles and blue triangles) is reduced. The non-monotonous behavior of the gate error in the original cycles (red squares) at smaller $R$ appears because in this regime the original cycle is so out-of-sync with the actual tilt angle $\tilde{\vartheta}$ that its corresponding rotations can accidently bring the qubit closer to the triplet state.

\begin{figure}[tbp]
\epsfxsize=0.95\columnwidth
\vspace*{0.0cm} \centerline{\epsffile{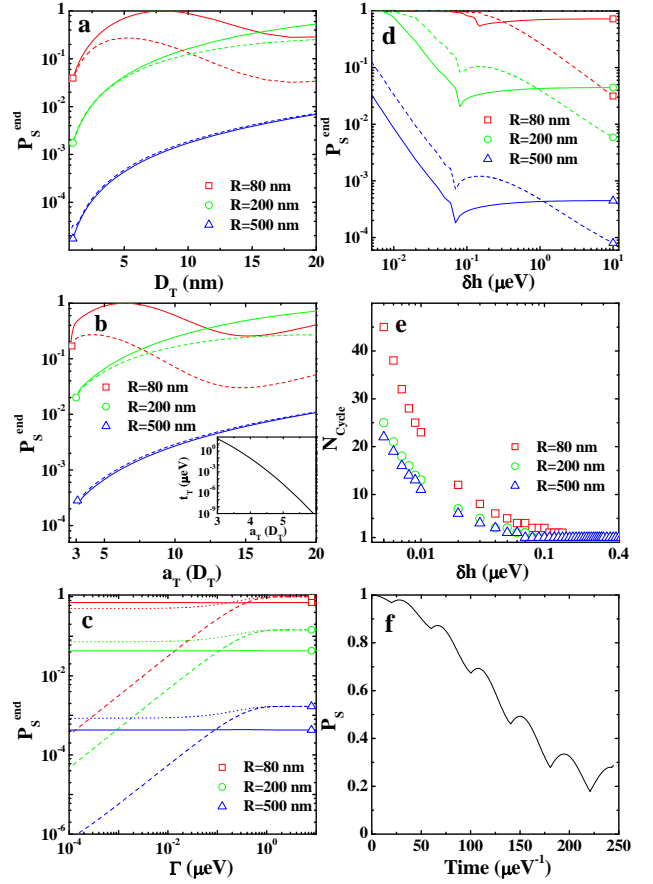}} \vspace{-0.2cm}
\caption{(Color online) $\pi$ flip error of $\beta-$corrected (solid lines) and $\beta'-$corrected (dashed lines) cycles for three qubit-TLS distances as a function of: (a) TLS Bohr radius $D_T$ ($a_T=4D_T$ at all points), (b) TLS centers half separation $a_T$ ($D_T=5$ nm for all points), (c) TLS spontaneous emission rate $\Gamma$ ($D_T=5$ nm, $a_T=20$ nm), and (d) magnetic field inhomogeneity between the two dots ($D_T=5$ nm, $a_T=20$ nm). The inset in (b) shows TLS tunneling vs. $a_T$. Dotted lines in (c) depict the gate error for the $\beta'-$corrected cycle with an initial TLS superposition state. (e) Number of cycles needed to complete the $\pi$ flip as a function of $\delta h$, for $\beta$-corrected cycles. (f) Singlet probability vs. time for $R=200$ nm, $\delta h=0.02 \mu$eV, where 7 $\beta-$corrected cycles are needed to complete the operation. In plots (a)-(c), the magnetic field inhomogeneity is $\delta h=1 \mu$eV and in all plots except (c), the TLS spontaneous emission rate is $\Gamma=0.1 \mu$eV.}
\label{Fig7n}
\end{figure}

The dependence of the 3-pulse $\pi$ flip gate errors on various system parameters is depicted in Figure \ref{Fig7n}. In order to isolate the effects of each parameter on gate errors, we fix all other parameters, unlike the calculation presented in figure \ref{Fig5n}, where $\Gamma,t_T,a_T$ were all varied consistently.\cite{Overdamp} Figures \ref{Fig7n}a-b show gate error dependence on the size of each of the TLS sites ($D_T$) and TLS half separation ($a_T$), respectively, for three qubit-TLS distances ($R=80, 200, 500$ nm). While $D_T$ is kept fixed in Figure \ref{Fig7n}b, $a_T$ is scaled with $D_T$ in figure \ref{Fig7n}a, thus the similar behavior of the gate error in both figures demonstrates that it is the distance between the TLS centers and not their size that impacts the gate error. This is consistent with our identification of the $\gamma$ coupling as the source of the gate error, since Eqs.~(\ref{gamma_dd})-(\ref{gamma_Qd}) show that $\gamma$ is proportional to $a_T$ and does not depend on $D_T$. At $R=200$ nm we obtain a gate error of 1\% for $a_T=175$ nm, and a 10\% error for $a_T=964$ nm for the $\beta-$corrected cycles. Notice that the difference in the remaining error in the two corrected cycles gets smaller as $R$ increases. The non-monotonous behavior observed in the $R=80$ nm case (red lines in Figs.~\ref{Fig7n}a and \ref{Fig7n}b) are due to higher-order contributions to $\beta$ in the multipole expansion ($\beta_{dQ},\beta_{QQ}$) that become significant at shorter ranges, and exhibit a complex dependence on $D_T$ (see Eqs.~(\ref{beta_dQ})-(\ref{beta_QQ})).

Figure \ref{Fig7n}c shows the gate error dependence on the TLS spontaneous emission rate $\Gamma$. The $\beta-$corrected cycles (solid lines) show very little dependence on $\Gamma$, similarly to the $X$ rotations studied in the preceding section. Since the cycle involves $\tilde{\vartheta}$ rotations for which $\Gamma$-dependence is present (see Figure \ref{Fig5n}d and related discussion), we conclude that the $\Gamma$-related effects of the first pulse in the cycle are erased by those of the third pulse, and the overall gate errors correspond to the $\gamma$ coupling. In contrast, the remaining error in the $\beta'-$corrected cycles (dashed lines), which corrects some of the $\gamma$-related effects, bring out the $\Gamma$ dependence. We find the $\beta'$ cycles error grows linearly with $\Gamma$: $Err_{\beta'}=a_1 \Gamma$ for all $R$ values, before saturating at a value corresponding to the $\gamma$ coupling. Saturation is reached at $\Gamma \approx 1\mu$eV, suggesting that the $\Gamma$ dynamics is governed by the TLS tunneling (which is the same for all $R$). For comparison we plot the gate error dependence on $\Gamma$ for the case of TLS initial state of equal superposition $\frac{1}{\sqrt{2}} \left( |\phi_L^T \rangle+|\phi_R^T \rangle \right)$ (dotted lines). This case presents a much weaker $\Gamma$ dependence since the TLS is set into its ground state, limiting the system dynamics. More work is needed to determine whether these $\Gamma$ dependencies are an artifact of our simplistic amplitude damping model for the TLS-charge environment coupling, or a generic feature characteristic to this system.

Figure \ref{Fig7n}d shows the gate error dependence on the qubit's magnetic field inhomogeneity $\delta h$. Below a certain $\delta h$ value ($0.15 \mu$eV for $R=80$ nm and $0.07 \mu$eV for $R=500$ nm), the $\beta$ corrected tilt angle $\tilde{\vartheta} > \pi/4$ and more cycles are needed to complete the $\pi$ flip, as seen in Fig.~\ref{Fig7n}e. In this regime the gate error increases for both corrections, obeying a power law corresponding to the increasing gate time. Above this threshold the gate errors of the $\beta'-$ corrected cycles (dashed lines) continue to scale with the gate time, whereas those of the $\beta-$corrected cycles (solid lines) are largely independent of $\delta h$, with a value corresponding to the $\gamma$ coupling (and thus to $R$). We find that the scaling of the entire sequence time with $\delta h$ obeys the power law $T_\pi=a_1 \delta h^{-a_2}$ with fitting parameters $a_1=3.23, a_2=1.08$ ($a_1=3.19, a_2=1.06$) for $R=80$ nm ($R=500$ nm), roughly corresponding to a gate time inversely proportional to $\delta h$ (slightly larger overhead is needed at shorter ranges due to the increased number of cycles). At sufficiently large $\delta h$ values error in the $Z$ rotation (2nd step in the cycle) is introduced. Since this error grows with $\delta h$ its effect competes with the $\gamma$ related errors, and may thus explain the error saturation for the $\beta-$corrected cycles. Apparently, the effect of the $Z$ axis error at large $\delta h$ is greatly reduced for the $\beta'-$ corrected cycles, where the $\tilde{\vartheta}$ tilt-angle includes the exchange-like contribution from $\gamma$, providing a more accurate rotation.

Figure \ref{Fig7n}f depicts the singlet probability for $\delta h=0.02 \mu$eV, $R=200$ nm, where 7 cycles are needed to complete the $\pi$ flip, in the $\beta$-corrected cycle (since $\vartheta>\pi/4$ in this case, the $\pi$ flip cannot be completed in the original cycle). We note that although the $\gamma$ coupling in this case is small ($\gamma_\vartheta =1$ neV), a large gate error of $28.5\%$ is found. This large error is mostly due to the small $\delta h$ value leading to an overall two-orders-of-magnitude increase in the operation time. We stress that for short qubit-TLS distances, the large $\beta$ coupling necessitates  an increasing number of cycles to complete the flip even when $\delta h$ is large. In this case the overall operation times are only slightly longer than a single-cycle gate and the resulting large gate-errors reflect the large $\gamma$ coupling and not the increasing number of cycles.

\subsection{Qubit rotation and dephasing at the $\tilde{J}=0$ sweet spot}
\label{Sweet}

\begin{figure}[tbp]
\epsfxsize=1.0\columnwidth
\vspace*{1.7cm} \centerline{\epsffile{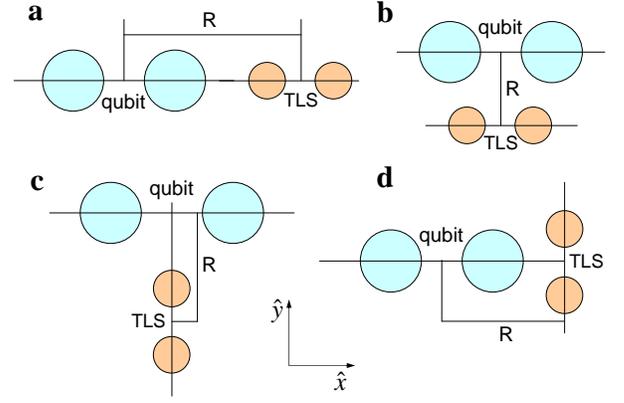}} \vspace{-2.9cm}
\caption{Several qubit-TLS geometries for $\theta=\theta_T=90^\circ$ (qubit and TLS lie in the $X-Y$ plane): (a) Same axis ($\hat{R}=\hat{x}_T=\hat{x}$); (b) Parallel axes ($\hat{R}=\hat{y}, \hat{x}_T=\hat{x}$); (c) Perpendicular axes ($\hat{R}=\hat{x}_T=\hat{y}$); (d) Perpendicular axes ($\hat{R}=\hat{x}, \hat{x}_T=\hat{y}$). For the geometries depicted in figures (a) and (b) the $\beta$ coupling is always positive, while for those in figures (c) and (d) it is always negative.}
\label{Fig8n}
\end{figure}

The form of the effective qubit exchange energy, $\tilde{J}=J-2\beta$ suggests that certain qubit-TLS geometries can yield bias points at which $\tilde{J}=0$. At such a bias one can perform $X$ rotations in the presence of $\delta h$ without resorting to either control over qubit tunneling or the three-pulse bias cycles described in the preceding section. In that sense a $\tilde{J}=0$ bias point is convenient for either performing single qubit rotations or doing nothing (idle position). It is thus conceivable that in an architecture based on double dot qubit, an additional nearby double dot could alleviate the quantum control of single qubit operations. We therefore consider a system consists of two double dots, one holding the encoded spin qubit and the other assisting in the qubit manipulation, and characterize the effects of the latter on the qubit at the $\tilde{J}=0$ working point. In general, the $\beta$ coupling can change its sign with $R$ but certain geometries yield same sign for $\beta(R)$ throughout the relevant distance range.\cite{beta0} For instance, vertically aligned double dots ($\theta=\phi_T=0, \theta_T=90^\circ$) yield $\beta<0$ for any $R$. Here we analyze geometries in which both subsystems lie in the same $X-Y$ plane. Figure \ref{Fig8n} shows four prototypical examples, where two configurations (Fig.~\ref{Fig8n}a-b) yield $\beta(R)>0$, and two (Fig.~\ref{Fig8n}c-d) yield $\beta(R)<0$ for all $R$. Sweet spots can thus be found for the two geometries depicted in Fig.~\ref{Fig8n}a-b.

\begin{figure}[tbp]
\epsfxsize=0.7\columnwidth
\vspace*{0.0cm} \centerline{\epsffile{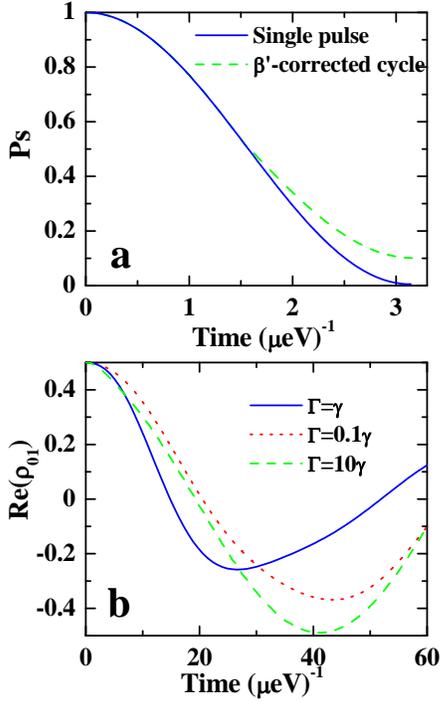}} \vspace{-0.1cm}
\caption{(Color online) Quantum control using a second double dot in same axis configuration (Fig.~\ref{Fig8n}a) with $R=200$ nm (a) A $\pi$ rotation about the $X$ axis with $\delta h=1 \mu$eV, obtained by a single pulse at the $\tilde{J}=0$ sweet spot, $\widetilde{\Delta x}=0.3866$ (solid-blue line) and by using a three-pulse cycle (dashed-green line). The two bias points in the three-pulse cycles are: $\widetilde{\Delta x}_\vartheta=0.357$, and $\widetilde{\Delta x}_z=0.414$. Here, $\Gamma=0.04\mu$eV. (b) Real part of the off-diagonal element of the qubit's reduced density matrix vs.~time at the sweet spot, for several values of $\Gamma$, with $\delta h=0$.
For this figure we used $D_T=10$ nm, $a_T=36$ nm corresponding to TLS tunneling rate of $t_T=0.38 \mu$eV.}
\label{Fig9n}
\end{figure}

To demonstrate quantum control using an ancillary double dot, we consider as an example the same axis configuration depicted in Fig.~\ref{Fig8n}a, with $R=200$ nm. For the second double dot (TLS) we take: $D_T=10$ nm, $a_T=36$ nm, $L_z^T=5$ nm, corresponding to TLS couple-tunneling $t_T=0.38 \mu$eV. The single-qubit gate is obtained using a single pulse at the sweet spot ($\widetilde{\Delta x}=0.3866$ for $R=200$ nm). Figure \ref{Fig9n}a shows the singlet probability vs.~time for a $\pi$-flip rotation with the qubit and TLS initial states taken, as before, to be a singlet and $|\phi^T_L \rangle$, respectively. The sweet spot gate operation (solid-blue line) presents a much smaller gate error (0.5\%) as compared with the equivalent gate operation obtained using the three-pulse $\beta'-$coorrected cycle discussed in the preceding section (dashed-green line; gate error 10\%). Notice that for the rotations shown in this figure the coupling terms are calculated with qubit-TLS relative orientations $(\theta, \phi, \theta_T, \phi_T)$ fixed by the specific system geometry and no angular averaging is employed.

Evidently, the $\tilde{J}=0$ bias point is a convenient idle working point. To study qubit dephasing at this point we consider $\delta h=0$ for which the qubit computational basis states, $S$, and $T_0$ are eigenstates of the Hamiltonian, Eq.~(\ref{H}). The qubit initial state is taken as the superposition $(|S \rangle +|T_0\rangle)/\sqrt{2}$, susceptible to dephasing (the TLS initial state is still $\phi^T_L \rangle$). Figure \ref{Fig9n}b shows the time dependence of the real part of $\rho_{01}^Q$, the off-diagonal element of the qubit's reduced density matrix, corresponding to the qubit's dephasing (the imaginary part of $\rho_{01}^{Q}$ exhibits a similar time scale). The time scale for dephasing is set by the $\gamma$ coupling, whereas the role played by the TLS-environment coupling $\Gamma$ is secondary. For $\Gamma \gg \gamma$ (dashed-green line) and $\Gamma \ll \gamma$ (dotted-red line), the $\gamma$ coupling, which entangles the qubit and TLS produces coherent phase oscillations. Only when $\gamma$ and $\Gamma$ are comparable in magnitude damping of the qubit phase oscillations occurs (solid-blue line in Fig.~\ref{Fig9n}b). These results are confirmed by the analytical solution of the master equation for the case of no tunneling between the TLS centers, given in appendix D. We stress that the TLS leads to fluctuations in the energy splitting between singlet and triplet states. The randomness in these fluctuations leading to qubit dephasing comes from the relaxation of the TLS to its reservoir rather than from direct interaction between the qubit and the reservoir.

It should be emphasized that the behavior described above is largely influenced by the choice of parameters, which fixes the location of the sweet spot. In particular, since the existence of a sweet spot requires sufficiently large $\beta$, in order to minimize gate errors and dephasing, one should look for system configurations that increase the $\beta/\gamma$ ratio.\cite{para} For the geometry considered here we find that gate errors and dephasing exhibit a non-monotonous dependence on $R$, due to the fact that the bias point at which $\tilde{J}=0$ becomes more negative as $R$ decreases (cl.~Fig.~\ref{Fig4n}). Since $\gamma$ increases for shorter $R$, but also decreases for more negative biases, these two opposing factors combine to give a non-monotonous behavior of the gate error. While the above discussion has demonstrated the existence of sweet spots that may aid in two-spin qubit manipulation, more work is needed in order to identify favorable working points, so that the proposed architecture becomes viable.

\section{Concluding remarks}
\label{conc}

In this paper we studied the effects of a nearby charge fluctuator (modeled as a TLS) on a qubit encoded by the two-electron spin states in a gate-defined double dot. We presented a quantitative analysis of the Coulomb coupling between the qubit orbital states and the TLS by means of a multipole expansion.
The resulting interaction terms were divided into the $\beta$ coupling ($I^T \times \sigma_z^Q$) that renormalizes the qubit exchange energy and the $\gamma$ coupling ($\sigma_z^T \times \sigma_z^Q$) that entangles the qubit and the TLS, and can therefore mediate decoherence effects to the spin qubit due to the charge environment. We find that $\gamma$ is generally smaller than $\beta$, and is comparable to the exchange energy at $R=100$ nm for the chosen system parameters. The ratio $\beta/\gamma$ is proportional to both the qubit-TLS distance $R$, and TLS centers separation $a_T$. We note that the $\beta/\gamma$ values stated in section \ref{3cycle} were obtained by averaging over all qubit-TLS orientations, and certain geometries will yield considerably different values.\cite{para} Both couplings depend strongly on $R$ and on the qubit bias. In particular, the couplings enhance considerably around the $(1,1)-(0,2)$ avoided crossing point, where the exchange energy bias dependence is strongest. Thus, we confirm as expected,\cite{Hu} that the spin qubit is most susceptible to charge-noise-induced decoherence when operated at positive detuning, at and above the anticrossing point.

As a first application of these results we employed a master equation formalism to study the spin decoherence effects due to the charge environment that are mediated by the nearby TLS.
We derived a set of differential equations for the density matrix describing the qubit-TLS system and solved it for various scenarios evaluating gate errors during single-qubit operations, and qubit dephasing times. We find competing dependence of the gate errors on qubit-TLS distance and orientations, leading to a non-monotonous behavior. For a single-pulse rotation, gate errors are found to be largest ($5-10\%$ for $\Gamma=0.04 \mu$eV) when qubit rotation is performed about an axis midway between the $X$ and $Z$ axes.
For positive biases, the $\beta$ coupling (at $R \lesssim 100$ nm) is large enough to produce a $Z$ rotation thus, although $\gamma$ is large in this regime, gate errors are very small.
These considerations demonstrate the need for a careful analysis of the system at hand in order to optimize its performance.

Analysis of a three-pulse cycle recently proposed to generate an $X$ rotation in the presence of finite exchange has indicated the implications of a nearby charge fluctuator on the feasibility of this scheme.
We find that the gate errors for a $\pi$-flip grow fast with $\gamma$, i.e., with decreasing qubit-TLS distance. For $R=500, 200, 80$ nm the gate errors are $0.04\%, 3.9\%, 66\%$ for the $\beta-$corrected cycle, and $0.02\%, 1.6\%, 12\%$ for the $\beta'-$corrected cycle, respectively, corresponding to $\gamma= 0.06, 0.88, 13.9$ neV at the $\tilde{\vartheta}$ working point. Moreover for $R \lesssim 30$ nm, with our parameter choice ($\delta h=1\mu$eV), the increase in the $\beta$ coupling produces a $\tilde{\vartheta}$-tilt rotation axis that approaches the $Z$ axis, thus more than one cycle are needed to complete the $\pi$ flip and the scheme becomes inefficient. The dependence of cycle errors on other system's parameters was also discussed.

Finally we identified certain qubit-TLS geometries for which a convenient working point exists such that the effective exchange energy vanishes. We analyze the qubit performance at this so-called sweet-spot in terms of gate errors and dephasing ($\delta h=0$, idle position), focusing on a system geometry in which the qubit and the TLS (provided here by an ancillary double dot) lie on the same axis. Our results suggest a possible qubit design that incorporates a double-dot qubit and an ancillary double-dot that serves to ease qubit manipulation by enabling working at the sweet spot. We find that gate errors and dephasing are sensitive to the system configuration, and in particular to the bias position of the sweet spot, thus a careful analysis is imperative to optimize qubit performance small gate errors. While this design entails an added complexity in calibration and initialization of the qubit, it provides an accessible and convenient working point for an idle position and single-qubit rotations.

In this paper we considered an indirect spin decoherence channel where a single TLS's relaxation to a reservoir is mediated to the qubit via their Coulomb-assisted entanglement.
Within this simple model, the role played by the TLS-reservoir coupling $\Gamma$ is secondary to that of $\gamma$, and we find that pure dephasing appears only when the two are comparable in magnitude.
Our study provides an initial step in the quantitative evaluation of the effects of charge environment on spin qubits, by microscopically calculating the qubit-TLS coupling. To make further progress, one may apply techniques developed in studies of superconducting Josephson qubits in the context of spin qubits. Notwithstanding the details pertaining to superconductor devices, fluctuating TLSs weakly coupled to the qubit were shown to produce both $1/f$ and Ohmic noise, inducing qubit relaxation and dephasing.\cite{Shnirman,Paladino,Bergli} Building on these ideas, one can evaluate decoherence effects for specific charge noise spectra mediated by the TLS. In addition one can consider also direct coupling of the qubit to the reservoir and evaluate whether the mediating TLS introduces a distinctively new decoherence channel.

To make contact with current experimental setups of gated QDs, one should identify possible candidates for charge fluctuators, in terms of their physical location and states. Two such mechanisms may be electrons jumping between two traps or between a localized state and a metallic gate.\cite{Paladino1,Faoro} We note that in order to produce a random telegraph noise, the energy splitting of the TLS should be smaller than $k_b T$ so that the switching rates for transitions between the two states $1 \rightarrow 2$ and $2 \rightarrow 1$ are comparable. Further measurements of the noise spectrum and its temperature dependence, similarly to those carried in superconducting devices\cite{Astafiev} will be instrumental in pointing at the correct mechanism.

\section*{Acknowledgments}

The authors acknowledge financial support by NSA/LPS through ARO.

\section*{Appendix A: qubit-TLS entanglement due to charge coupling}
\renewcommand{\theequation}{A-\arabic{equation}}
\setcounter{equation}{0}

Here we evaluate the concurrence of the qubit-TLS system state evolving under the interaction Hamiltonian, Eq.~(\ref{Hint}):
\begin{equation}
|\psi (t) \rangle = e^{-i{\cal H}_{\rm int} t} |\psi_T(t=0) \rangle \otimes |\psi_Q(t=0) \rangle
\label{psit}
\end{equation}
We find that
\begin{eqnarray}
e^{-i {\cal H}_{\rm int} t} \!&\!=\!&\!\zeta_0(t)(I^T\otimes I^Q) +\zeta_1(t) (I^T\otimes \sigma_z^Q) \nonumber \\
\!&\!+\!&\! \zeta_2(t)(\sigma_z^T \otimes I^Q)+\zeta_3(t)(\sigma_z^T\otimes \sigma_z^Q)
\end{eqnarray}
with
\begin{eqnarray}
\zeta_0(t)\!&\!=\!&\!\cos \alpha t \cos \beta t \cos \gamma t +i \sin \alpha t \sin \beta t \sin \gamma t \nonumber \\
\zeta_1(t)\!&\!=\!&\!\sin \alpha t \cos \beta t \sin \gamma t +i \cos \alpha t \sin \beta t \cos \gamma t \nonumber \\
\zeta_2(t)\!&\!=\!&\!\cos \alpha t \sin \beta t \sin \gamma t +i \sin \alpha t \cos \beta t \cos \gamma t \nonumber \\
\zeta_3(t)\!&\!=\!&\!-\sin \alpha t \sin \beta t \cos \gamma t -i \cos \alpha t \cos \beta t \sin \gamma t \nonumber.
\end{eqnarray}
The state Eq.~(\ref{psit}) is now expanded in the magic basis $\psi(t)=\sum_i \alpha_i |e_i \rangle$ where\cite{HilWoo}
\begin{eqnarray*}
|e_1 \rangle \!&\!=\!&\! \frac{1}{2} \left( | \uparrow \uparrow\rangle +|\downarrow \downarrow \rangle \right) \\
|e_2 \rangle \!&\!=\!&\! \frac{i}{2} \left( | \uparrow \uparrow\rangle -|\downarrow \downarrow \rangle \right)\\
|e_3 \rangle \!&\!=\!&\! \frac{i}{2} \left( | \uparrow \downarrow\rangle +|\downarrow \uparrow \rangle \right)\\
|e_4 \rangle \!&\!=\!&\! \frac{1}{2} \left( | \uparrow \downarrow\rangle -|\downarrow \uparrow \rangle \right).
\end{eqnarray*}
Here the left (right) pseudospin corresponds to the TLS (qubit) state. The concurrence is given by $C(\psi)=\left|\sum_i \alpha_i^2\right|$ and provides an entanglement measure between the two subsystems.
Taking the initial states of the two subsystems as a (normalized) superposition of their basis states:
\begin{eqnarray*}
|\psi_T(0)\rangle\!&\!=\!&\! a_L |L\rangle +a_R |R\rangle \\
|\psi_Q(0)\rangle\!&\!=\!&\! a_S |S\rangle +a_T |T\rangle
\end{eqnarray*}
we find the expansion coefficients as:
\begin{eqnarray*}
\alpha_1(t) \!&\!=\!&\! a_S a_L (\zeta_0\!-\!\zeta_1\!-\!\zeta_2\!+\!\zeta_3)\!+\!a_T a_R(\zeta_0\!+\!\zeta_1\!+\!\zeta_2\!+\!\zeta_3) \\
\alpha_2(t) \!&\!=\!&\! ia_S a_L (\zeta_0\!-\!\zeta_1\!-\!\zeta_2\!+\!\zeta_3)\!-\!ia_T a_R(\zeta_0\!+\!\zeta_1\!+\!\zeta_2\!+\!\zeta_3) \\
\alpha_3(t) \!&\!=\!&\! -ia_S a_L (\zeta_0\!-\!\zeta_1\!+\!\zeta_2\!-\!\zeta_3)\!-\!ia_T a_R(\zeta_0\!+\!\zeta_1\!-\!\zeta_2\!-\!\zeta_3) \\
\alpha_4(t) \!&\!=\!&\! -a_S a_L (\zeta_0\!-\!\zeta_1\!+\!\zeta_2\!-\!\zeta_3)\!+\!a_T a_R(\zeta_0\!+\!\zeta_1\!-\!\zeta_2\!-\!\zeta_3) \\
\end{eqnarray*}
and the concurrence takes the simple form:
\begin{eqnarray}
C(\psi)\!&\!=\!&\! \left|\sum_i \alpha_i^2 \right|=|16a_La_Ra_Sa_T(\zeta_0 \zeta_3 -\zeta_1 \zeta_2 )| \nonumber \\
\!&\!=\!&\! 8a_La_Ra_Sa_T |\sin(2\gamma t)|.
\end{eqnarray}
Thus, the qubit and the TLS are entangled only in the presence of the $\gamma$ coupling and when their individual states are in a superposition of their basis states.

\section*{Appendix B: Orbital Hamiltonian for a biased double-dot}
\renewcommand{\theequation}{B-\arabic{equation}}
\setcounter{equation}{0}

In this appendix we provide details of the calculation of the orbital Hamiltonian, Eq.~(\ref{Horb}) within the Hund-Mulliken framework. This work extends the results of Burkard and Loss\cite{BurLos} to a biased dot configuration.
Taking the $X$-axis ($Z$-axis) along the electric (magnetic) field, the two-electron orbital Hamiltonian is given by
\begin{equation}
H_{\rm orb}=\sum_{i=1,2} H_i^{\rm SP}+C({\bf r}_1,{\bf r}_2)
\end{equation}
where $C({\bf r}_1,{\bf r}_2)=e^2/\kappa |{\bf r}_1-{\bf r}_2|$ is the Coulomb interaction between the two electrons, and the single-particle Hamiltonian is
\begin{equation}
H_i^{\rm SP}=\frac{1}{2m} \left( {\bf p}_i-\frac{e}{c} {\bf A} ({\bf r}_i) \right)^2 +ex_i E+V({\bf r}_i).
\end{equation}
The double-dot confinement potential is modeled using a quartic potential in the $x-y$ plane and a finite potential in the $z$ direction
\begin{eqnarray}
V({\bf r}_i)&=&V_{xy}(x,y)V_z(z) \nonumber \\
V_{xy}(x,y)&=& \frac{m \omega_0^2}{2} \left[ \frac{1}{4a^2}(x^2-a^2)^2+y^2 \right] \\
V_z(z)&=& \left\{ \begin{array}{ll} 0 & |z| \leq L_z/2 \\ V_z & |z|>L_z/2 \end{array} \right. \nonumber
\end{eqnarray}
where we consider a much stronger confinement in the $z$ direction, appropriate for typical gate-defines QD structures. This enables us to perform separation of variables in the lateral and $z$ directions, and to approximate the Coulomb interactions using 2-D integrals. The matrix elements of the orbital Hamiltonian are found by adding and subtracting the harmonic potentials centered at $\pm a$,\cite{BurLos} thus we have
\begin{eqnarray}
H_{\rm orb}\!&\!=\!&\! h_{-a}({\bf r}_1)+h_{a}({\bf r}_2)+W_-({\bf r}_1)+W_+({\bf r}_2)+C \nonumber \\
h_{\pm a}({\bf r}) \!&\!=\!&\! \frac{1}{2m} \left( {\bf p}-\frac{e}{c} {\bf A} ({\bf r}) \right)^2 \nonumber \\
\!&\!+\!&\! \frac{m \omega_0^2}{2} \left[ (x\mp a)^2+y^2 \right] +eEx \\
W_\pm ({\bf r}) \!&\!=\!&\! \frac{m \omega_0^2}{2} \left[ \frac{x^4}{4a^2}-\frac{3x^2}{2}-\frac{3a^2}{4} \pm 2xa \right] \nonumber
\end{eqnarray}
Using the orthonormalized single-particle orbitals, $\psi_{\pm a}={\cal N}(\phi_{\pm a} -g \phi_{ \mp a})$ we find the single-particle energies and tunnelings in $\hbar \omega_0$ units:
\begin{eqnarray}
\epsilon_\pm &=&\langle \psi_{\pm a} |h_{\pm a}+W_\pm | \psi_{\pm a} \rangle = \epsilon^{0}+\epsilon_1^E \pm\epsilon_2^E \nonumber \\
\tilde{\epsilon}_\pm &=&\langle \psi_{\mp a} |h_{\pm a}+W_\pm | \psi_{\mp a} \rangle = \epsilon^{0}+\epsilon_1^E \mp \epsilon_2^E \\
t &=&\langle \psi_{\pm a} |h_{\pm a}^0+W_\pm | \psi_{\mp a} \rangle = t^{0}+t^E \nonumber
\end{eqnarray}
with
\begin{eqnarray}
\epsilon^{0}&=& b+\frac{3}{32b^2d^2}+\frac{3}{8}\frac{s^2}{1-s^2} \left(\frac{1}{b}+d^2\right) \nonumber \\
\epsilon_1^E&=& \widetilde{\Delta x}^2\left( \frac{5s^2-2}{4(1-s^2)}+\frac{3}{8bd^2}+\frac{\widetilde{\Delta x}^2}{8d^2}\right) \nonumber  \\
\epsilon_2^E&=& \frac{\widetilde{\Delta x}d(1-g^2)}{1-2sg+g^2}\left(1-\frac{3}{4bd^2}-\frac{\widetilde{\Delta x}^2}{2d^2}\right) \\
t^{0}&=&-\frac{3s}{8(1-s^2)} \left(\frac{1}{b}+d^2\right) \nonumber \\
t^E&=&-\frac{3s\widetilde{\Delta x}^2}{4(1-s^2)}. \nonumber
\end{eqnarray}
Here $b$ is the magnetic compression factor, $d,\widetilde{\Delta x}$ are the dot half-separation and electric-field-induced orbital shift in units of Bohr radius, respectively, and $s$ is the wavefunction overlap, given in the main text.
Notice that the zero-bias results $\epsilon^{0}$, and $t^0$ coincide with ref.~\onlinecite{BurLos}.
In the basis of the three singlet and separated triplet states, the matrix elements of the orbital Hamiltonian Eq.~(\ref{Horb}) are found as
\begin{eqnarray}
\epsilon_{20}^S&=& \langle \psi_{-a} \psi_{-a} | H_{\rm orb} | \psi_{-a} \psi_{-a} \rangle = \epsilon_- +\tilde{\epsilon}_+ +U \nonumber \\
\epsilon_{02}^S&=& \langle \psi_{a} \psi_{a} | H_{\rm orb} | \psi_{a} \psi_{a} \rangle = \epsilon_+ +\tilde{\epsilon}_- +U \nonumber  \\
\epsilon_{11}^{S/T}&=& \frac{1}{2} \langle \psi_{a} \psi_{-a} \pm \psi_{-a} \psi_a | H_{\rm orb} | \psi_{a} \psi_{-a} \pm \psi_{-a} \psi_a \rangle \nonumber  \\
&=& \frac{1}{2}(\epsilon_+ +\epsilon_- +\tilde{\epsilon}_+ +\tilde{\epsilon}_-) +V_\pm  \\
t_H&=& -\frac{1}{2} \langle \psi_{\pm a} \psi_{\pm a} | H_{\rm orb} | \psi_a \psi_{-a} + \psi_{-a} \psi_{a} \rangle=-t-T, \nonumber
\end{eqnarray}
where the upper (lower) sign in $\epsilon_{11}^{S/T}$ corresponds to the separated singlet (triplet) energy. The two-electron Coulomb matrix elements are
\begin{eqnarray}
U &=& \langle \psi_{\pm a} \psi_{\pm a} |C | \psi_{\pm a} \psi_{\pm a} \rangle \nonumber \\
V_{\pm} &=& \frac{1}{2} \langle \psi_a \psi_{-a}\pm \psi_{-a} \psi_a |C | \psi_{a}\psi_{-a} \pm \psi_{-a}\psi_a \rangle \nonumber \\
T &=& \frac{1}{2} \langle \psi_a \psi_{-a}+\psi_{-a} \psi_a |C | \psi_{\pm a} \psi_{\pm a} \rangle \\
X &=& \frac{1}{2} \langle \psi_{\pm a} \psi_{\pm a} |C | \psi_{\mp a} \psi_{\mp a} \rangle \nonumber
\end{eqnarray}
and their closed-form expressions can be found in ref.~\onlinecite{BurLos}.\cite{F4}

\section*{Appendix C: qubit-TLS Coulomb coupling terms}
\renewcommand{\theequation}{C-\arabic{equation}}
\setcounter{equation}{0}
The qubit-TLS Coulomb coupling terms in Eqs.~(\ref{alpha})-(\ref{gamma}) were calculated to quadrupole-quadruploe order. To 2nd order in the qubit orbital overlap, $s$, they are given by the following expressions
\begin{widetext}
\begin{eqnarray}
\alpha_{qd} \!&\!=\!&\! \frac{2e^2\tilde{a}_T a_B }{\varepsilon R^2} \left[ \sin \theta \sin \theta_T \cos(\phi-\phi_T)+\cos \theta \cos \theta_T \right] \label{alpha_qd} \\
\alpha_{dd} \!&\!=\!&\! -\frac{e^2 \tilde{a}_T a_B^2}{\varepsilon R^3} \left[2\widetilde{\Delta x}+d{\cal N}_S^2 (a_1^2-a_2^2) \right] \left[\sin \theta_T \cos \phi_T -3\sin \theta \cos \phi \left( \sin \theta \sin \theta_T \cos(\phi-\phi_T)+\cos \theta \cos \theta_T \right) \right] \\
\alpha_{Qd} \!&\!=\!&\! \frac{e^2 \tilde{a}_T a_B^3}{\varepsilon R^4 b} \left\{ -2 \left[1+b(\widetilde{\Delta x}^2+d^2-2 \tilde{l}_z^2 ) \right]\cos \theta \cos \theta_T+\left[1+2b (2\widetilde{\Delta x}^2+2d^2-\tilde{l}_z^2) \right] \notag \cos \phi \cos\phi_T \sin \theta \sin \theta_T \right. \nonumber \\
\!&\!+\!&\! \left. \left[1-2b(\widetilde{\Delta x}^2+d^2+ \tilde{l}_z^2) \right]\sin \phi \sin\phi_T \sin \theta \sin \theta_T + \frac{5}{2} \left[1+b(\widetilde{\Delta x}^2+d^2-2\tilde{l}_z^2)(2\cos^2 \theta-\sin^2 \theta) \right. \right. \nonumber \\
\!&\!-\!&\! \left. \left. 3b(\widetilde{\Delta x}^2+d^2) \cos 2\phi \sin^2 \theta \right] \left(\cos\theta \cos \theta_T+\cos(\phi-\phi_T) \sin \theta \sin \theta_T \right) \right\}-\gamma_{Qd}
\end{eqnarray}
\begin{eqnarray}
\beta_{dq} \!&\!=\!&\! \frac{e^2 d a_B}{\varepsilon R^2} {\cal N}_S^2 (a_1^2-a_2^2) \sin \theta \cos \phi \label{beta_dq}\\
\beta_{Qq}\!&\!=\!&\! -\frac{e^2 d a_B^2 {\cal N}_S^2}{\varepsilon R^3}\left\{ sd \frac{a_1+a_2}{\sqrt{2}} \left[ (3\cos^2 \theta-1)+\frac{1}{b^2} (3\sin^2 \theta \sin^2 \phi-1) \right]
\!+\!\widetilde{\Delta x} (a_1^2-a_2^2) \left(3 \sin^2 \theta \cos^2 \phi-\! 1 \right) \right\}\\
\beta_{dQ} \!&\!=\!&\! -\frac{3e^2d a_B^3 {\cal N}_S^2}{8 \varepsilon R^4} (a_1^2-a_2^2) \sin \theta \cos \phi \left[ \cos 2 \phi_T \left(\tilde{a}^2_T+\tilde{D}_T^2-2\tilde{l}_{zT}^2-\tilde{a}_T^2 \cos 2\theta_T \right) (5\cos 2\phi \sin^2 \theta-2)\right. \notag \\
\!&\!+\!&\! \left. \left(\tilde{a}^2_T+\tilde{D}_T^2-2\tilde{l}_{zT}^2+3\tilde{a}_T^2 \cos 2\theta_T \right)(5 \cos^2 \theta -1) \right]
\label{beta_dQ} \\
\beta_{QQ} \!&\!=\!&\! \frac{3e^2d a_B^4 {\cal N}_S^2}{32 \varepsilon R^5} \left\{ \frac{\sqrt{2}ds(a_1+a_2)}{b^2} \left[ \left(\tilde{a}^2_T+\tilde{D}_T^2-2\tilde{l}_{zT}^2+3\tilde{a}_T^2 \cos 2\theta_T \right) \left( (1-2b^2)(35 \cos^4 \theta -30 \cos^2 \theta+3)\right. \right. \right.\notag \\
\!&\!+\!&\! \left. \left. \left. 5\cos 2\phi (7 \cos^2 \theta-1)\sin^2 \theta \right) + \cos 2\phi_T \left(\tilde{a}^2_T+\tilde{D}_T^2-2\tilde{l}_{zT}^2-\tilde{a}_T^2 \cos 2\theta_T \right) \left( 20 \cos^2 \theta -16+35 \sin^4 \theta \cos^2 2\phi \right. \right. \right. \notag \\
\!&\!+\!&\! \left. \left. \left. 5(1-2b^2) \cos 2\phi (7 \cos^2 \theta-1) \sin^2 \theta \right) \right] -\widetilde{\Delta x}(a_1^2-a_2^2) \left[ (3-30 \cos^2 \theta +35 \cos^4 \theta) \left( (\tilde{a}^2_T+\tilde{D}_T^2-2\tilde{l}_{zT}^2) \right. \right. \right. \notag \\
\!&\! \times \!&\!  \left. \left. \left. (-2+\cos 2\phi_T)-\tilde{a}_T^2 (6+ \cos 2 \phi_T) \cos 2\theta_T \right)+5 \cos 2\phi (5+7 \cos 2\theta) \left( \tilde{a}_T^2(3+\cos 2\phi_T) \cos 2\theta_T \right. \right. \right. \notag \\ \!&\!+\!&\! \left. \left. \left. 2(\tilde{a}^2_T+\tilde{D}_T^2-2\tilde{l}_{zT}^2) \sin^2 \phi_T \right) \sin^2 \theta +35 \cos 4 \phi \cos 2 \phi_T (\tilde{a}^2_T+\tilde{D}_T^2-2\tilde{l}_{zT}^2-\tilde{a}_T^2 \cos 2\theta_T ) \sin^4 \theta \right] \right\}
\label{beta_QQ}
\end{eqnarray}

\begin{eqnarray}
\gamma_{dd} \!&\!=\!&\! \frac{e^2 \tilde{a}_T d a_B^2 {\cal N}_S^2}{\varepsilon R^3} (a_1^2-a_2^2) \left[\sin \theta_T \cos \phi_T -3\sin \theta \cos \phi \left( \sin \theta \sin \theta_T \cos(\phi-\phi_T)+\cos \theta \cos \theta_T \right) \right] \label{gamma_dd} \\
\gamma_{Qd} \!&\!=\!&\! \frac{2e^2 \tilde{a}_T d a_B^3 {\cal N}_S^2}{\varepsilon R^4} \left\{sd\frac{a_1+a_2}{\sqrt{2}} \left[ \left(1+\frac{1}{b^2} \right) \cos \phi \cos \phi_T \sin \theta \sin \theta_T -\left( \frac{2}{b^2}-1 \right) \sin \phi \sin \phi_T \sin \theta \sin \theta_T \right. \right.\notag \\
\!&\!-\!&\! \left. \left. \left(2-\frac{1}{b^2} \right) \cos \theta \cos \theta_T+\frac{5}{2} \left(\cos \theta \cos \theta_T +\cos(\phi-\phi_T) \sin \theta \sin \theta_T \right) \left(-\left(1+\frac{1}{b^2} \right) \cos^2 \phi \sin^2 \theta \right. \right.\right. \notag \\
\!&\!+\!&\! \left. \left. \left. \left( \frac{2}{b^2}-1 \right) \sin^2 \phi \sin^2 \theta + \left(2-\frac{1}{b^2} \right) \cos^2 \theta \right) \right] - \frac{\widetilde{\Delta x}}{2}(a_1^2-a_2^2) \left[ 3 \cos \theta \cos \theta_T (1-5\cos^2 \phi \sin^2 \theta ) \right. \right. \nonumber \\
\!&\!+\!&\! \left. \left. \sin \theta \sin \theta_T \left(4 \cos \phi \cos \phi_T - 2 \sin \phi \sin \phi_T +5 \cos(\phi-\phi_T)(1-3\cos^2 \phi \sin^2 \theta )\right)\right]\right\}.
\label{gamma_Qd}
\end{eqnarray}
\end{widetext}
\section*{Appendix D: Analytic solution for the qubit-TLS-reservoir master equation for the case of zero TLS tunneling}
\renewcommand{\theequation}{D-\arabic{equation}}
\setcounter{equation}{0}

In this appendix we present an analytic solution to the master equation (\ref{rhot}) for the case of zero TLS tunneling, $t_T=0$.
The $\tilde{\sigma}_\pm$ operators in the interaction picture, Eq.~(\ref{sigma}) reduce in this case to
\begin{eqnarray}
\tilde{\sigma}_\pm \!&\!=\!&\! e^{\pm 2i(\omega_T -\alpha)} \left[\cos 2\gamma t (\sigma_\pm^T \otimes I^Q) \right. \nonumber \\
\!&\!\pm \!&\! \left. i\sin 2\gamma t (\sigma_\pm^T \otimes \sigma_z^Q) \right].
\end{eqnarray}
The differential equations for the system density matrix assume a simple form:
\begin{equation}
\frac{1}{\Gamma} \dot{ \tilde{\rho}} = \left(
\begin{array}{llll}
-2 \tilde{\rho}_{00} & -2\tilde{\rho}_{01} & -\tilde{\rho}_{02}                & -\tilde{\rho}_{03} \\
-2 \tilde{\rho}_{10} & -2\tilde{\rho}_{11} & -\tilde{\rho}_{12}                & -\tilde{\rho}_{13} \\
-2 \tilde{\rho}_{20} & -\tilde{\rho}_{21}  & 2\tilde{\rho}_{00}                & 2\tilde{\rho}_{01} e^{-4i \gamma t} \\
-2 \tilde{\rho}_{30} & -\tilde{\rho}_{31}  & 2\tilde{\rho}_{10}e^{4i \gamma t} & 2\tilde{\rho}_{11}
\end{array}
\right),
\label{dGamma}
\end{equation}
and are easily solved, with the appropriate initial conditions. In order to transform back to $\rho (t)$, we calculate $e^{-i {\cal H}t}$, where the system Hamiltonian is given in Eq.~(\ref{H}) with
${\bf B}_Q=\frac{1}{2} (\delta h,0,J)$, and ${\bf B}_T=(0,0,\omega_T)$. The resulting matrix is block diagonal:
\begin{widetext}
\begin{equation}
e^{-i {\cal H} t} = \left(
\begin{array}{cccc}
\cos \eta_- t+i\frac{B_z^-}{\eta_-} \sin \eta_- t & -i\frac{B_{Qx}}{\eta_-} \sin \eta_- t & 0 & 0 \\
-i\frac{B_{Qx}}{\eta_-} \sin \eta_- t & \cos \eta_- t-i\frac{B_z^-}{\eta_-} \sin \eta_- t & 0 & 0 \\
0 & 0 & \cos \eta_+ t+i\frac{B_z^+}{\eta_+} \sin \eta_+ t & -i\frac{B_{Qx}}{\eta_+} \sin \eta_+ t \\
0 & 0 & -i\frac{B_{Qx}}{\eta_+} \sin \eta_+ t & \cos \eta_+ t-i\frac{B_z^+}{\eta_+} \sin \eta_+ t
\end{array}
\right)
\label{expiHt}
\end{equation}
\end{widetext}
where we have defined $B_{z}^\pm = -B_{Qz} +\beta \pm \gamma$, and $\eta_\pm=\sqrt{B_{Qx}^2+B_z^{\pm2}}$.
The qubit dynamics can then be examined by letting the system evolve under the above Hamiltonian, and tracing out the TLS subsystem, thereby obtaining the qubit reduced density matrix.

Eqs.~(\ref{dGamma})-(\ref{expiHt}) are used to evaluate the gate error of a single-qubit rotation due charge coupling with TLS-environment amplitude damping.
We perform a rotation about the $\vartheta$-tilted axis, where $\vartheta =\arctan (J/\delta h)$, thus the qubit pseudofield vector lies in the $X-Z$ plane.
Taking the initial qubit state as a singlet, and the TLS initial state as an equal superposition of L/R states:
\begin{equation}
\psi_Q (t=0)=  \left(
\begin{array}{c} 1 \\ 0 \end{array} \right); \hspace{0.5 cm}
\psi_T (t=0)= \frac{1}{\sqrt{2}} \left(
\begin{array}{c} 1 \\ 1 \end{array} \right)
\end{equation}
the singlet probability as a function of time is found to be
\begin{eqnarray}
P_S(t)\!&\!=\!&\! \rho_{00}(t)+\rho_{22}(t)=1-B^2_{Qx}\left(\frac{\sin \eta_+ t }{\eta_+} \right)^2 \nonumber \\
\!&\!-\!&\! \frac{B_{Qx}^2}{2} e^{-2 \Gamma t}\! \left[ \!\left( \frac{\sin \eta_- t}{\eta_-}\right)^2  \!\! - \! \left( \frac{\sin \eta_+ t}{\eta_+}\right)^2  \right]. \label{Pst}
\end{eqnarray}
Without the qubit-TLS coupling, the singlet probability at the end of a $\pi$-rotation about the $\vartheta$-tilted axis ($T_\pi =\pi/2B_Q$) is
\begin{equation}
P_S(T_\pi)=1-\cos^2 \vartheta.
\end{equation}
To evaluate the pulse error due to the qubit-TLS coupling we design our new $\pi$-pulse to match the renormalized qubit exchange energy: $J \rightarrow \tilde{J}=J-2\beta$. Thus we have an effective pseudospin field $\tilde{{\bf B}}=\frac{1}{2}(\delta h,0,J-2\beta)$, which redefines the tilting angle to $\tilde{\vartheta}=(J-2\beta)/\tilde{B}$, and the corresponding the $\pi$-pulse time to $\tilde{T}_\pi =\pi/2\tilde{B}$. With the effect of the $\beta$ coupling removed the remaining pulse error is due to $\gamma$, and $\Gamma$. To second order in $\gamma/\tilde{B}$ Eq.~(\ref{Pst}) yields
\begin{eqnarray}
P_S(\tilde{T}_\pi)\!&\!=\!&\!1-\cos^2 \tilde{\vartheta} + \sin 2\tilde{\vartheta} \cos \tilde{\vartheta}\left[ 1 - e^{-\pi \Gamma/\tilde{B}} \right]\frac{\gamma}{\tilde{B}}\nonumber \\
\!&\!+\!&\! \cos^2 \tilde{\vartheta} \left[ 1- \left( 4-\frac{\pi^2}{4} \right) \sin^2 \tilde{\vartheta} \right] \left(\frac{\gamma}{\tilde{B}} \right)^2.
\label{P_Sanal}
\end{eqnarray}
Inspecting Eq.~(\ref{P_Sanal}) we find that the leading term in the gate error vanishes for rotations about the $X$ axis ($\tilde{\vartheta}=0$) or about the $Z$ axis ($\tilde{\vartheta}=\pi/2$). Thus the error we expect to observe in a $\pi$ flip operation will result, predominantly, from the 2nd order term, which is independent of $\Gamma$, the latter entering only odd order terms in the $\gamma/\tilde{B}$ expansion. The $\Gamma$ decay contribution will thus be effective only for rotations about a $\theta$ tilted axis. It should be noted that the convergence of Eq.~(\ref{P_Sanal}) depends on the ratio $\beta/\gamma$, thus, as the TLS size $a_T$ increases, $\beta/\gamma$ becomes smaller, and more terms are needed in the $\gamma/\tilde{B}$ expansion.

Next we use Eqs.~(\ref{dGamma})-(\ref{expiHt}) to evaluate qubit dephasing at the sweet spot $\tilde{J}=0$ (see section \ref{Sweet}). In order to have an idle working point we take $\delta h=0$ so that $\tilde{\bf B}=0$. We consider an equal superposition of singlet and triplet (L and R) states as our initial qubit (TLS) state. In this case $\eta_\pm =\gamma$ and the equations are trivial with the following solution for the off diagonal element of the qubit reduced density matrix:
\begin{eqnarray}
\rho_{01}^Q(t)\!&\!=\!&\!\rho_{01}(t)+\rho_{23}(t)=\frac{1}{2} e^{2i\gamma t} -\frac{i\gamma}{\Gamma +2i\gamma} \times \nonumber \\
&&\left[ \frac{1}{2} e^{-2i\gamma t} \left(1-e^{-2 \Gamma t} \right)+i\sin 2\gamma t\right]
\end{eqnarray}
The asymptotic behavior of this equation is:
\begin{equation}
\begin{array}{ll}
\rho_{01}^Q(t) \approx \frac{1}{2} \cos 2\gamma t, & \Gamma \ll \gamma \\
\rho_{01}^Q(t) \approx \frac{1}{2} e^{2i\gamma t} \left( 1-\frac{i \gamma}{\Gamma} \right), & \Gamma \gg \gamma.
\end{array}
\end{equation}
Thus, for our simple amplitude-damping model we find that the TLS-environment coupling $\Gamma$ has an impact on the qubit dephasing only when $\Gamma \approx \gamma$.
This behavior is demonstrated in Figure \ref{Fig9n}b.

\bibliographystyle{amsplain}

\begin{thebibliography}{99}
%
\bibitem{HanRMP} R.~Hanson, L.~P.~Kouwenhoven, J.~R.~Petta, S.~Tarucha, and L.~M.~K. Vandersypen, Rev.~Mod.~Phys.~{\bf 79}, 1217 (2007).
\bibitem{Ciorga} M.~Ciorga, A.~S.~Sachrajda, P.~Hawrylak, C.~Gould, P.~Zawadzki, S.~Jullian, Y.~Feng, and Z.~Wasilewski, Phys.~Rev.~B {\bf 61}, R16315 (2000).
\bibitem{Hanson} R.~Hanson, L.~H.~Willems van Beveren, I.~T.~Vink, J.~M.~Elzerman, W.~J.~M.~Naber, F.~H.~L.~Koppens, L.~P.~Kouwenhoven, and L.~M.~K.~Vandersypen, Phys.~Rev.~Lett.~{\bf 94}, 196802 (2005).
\bibitem{Petta} J.~R.~Petta, A.~C.~Johnson, J.~M.~Taylor, E.~A.~Laird, A.~Yacoby, M.~D.~Lukin, C.~M.~Marcus, M.~P.~Hanson, A.~C.~Gossard, Science {\bf 309}, 2180 (2005).
\bibitem{Amasha} S.~Amasha, K.~MacLean, Iuliana P.~Radu, D.~M.~Zumbuhl, M.~A.~Kastner, M.~P.~Hanson, and A.~C.~Gossard, Phys.~Rev.~Lett.~{\bf 100}, 046803 (2008).
\bibitem{YacobyPC} Amir Yacoby, private communication.
\bibitem{Ono} K.~Ono, D.~G.~Austing, Y.~Tokura, and S.~Tarucha, Science {\bf 297}, 1313 (2002).
\bibitem{Elzerman} J.~M.~Elzerman, R.~Hanson, L.~H.~Willems van Beveren, B.~Witkamp, L.~M.~K.~Vandersypen, and L.~P.~Kouwenhoven, Nature {\bf 430}, 431 (2004).
\bibitem{DiCarlo} L.~DiCarlo, H.~J.~Lynch, A.~C.~Johnson, L.~I.~Childress, K.~Crockett, C.~M.~Marcus, M.~P.~Hanson, and A.~C.~Gossard, Phys.~Rev.~Lett.~{\bf 92}, 226801 (2004).
\bibitem{Koppens} F.~H.~L.~Koppens, C.~Buizert, K.~J.~Tielrooij, I.~T.~Vink, K.~C.~Nowack, T.~Meunier, L.~P.~Kouwenhoven, and L.~M.~K.~Vandersypen, Nature {\bf 442}, 766 (2006).
\bibitem{Nowack} K.~C.~Nowack, F.~H.~L.~Koppens, Yu.~V.~Nazarov, and L.~M.~K.~Vandersypen, Science {\bf 318}, 1430 (2007).
\bibitem{Golovach} V.~N.~Golovach, A.~V.~Khaetskii, and D.~Loss, Phys.~Rev.~Lett.~{\bf 93}, 016601 (2004).
\bibitem{Merkulov} I.~A.~Merkulov, Al.~L.~Efros, and M.~Rosen, Phys.~Rev.~B {\bf 65}, 205309 (2002).
\bibitem{Johnson} A.~C.~Johnson, J.~R.~Petta, J.~M.~Taylor, A.~Yacoby, M.~D.~Lukin, C.~M.~Marcus, M.~P.~Hanson, and A.~C.~Gossard, Nature {\bf 435}, 925 (2005).
\bibitem{Koppens1} F.~H.~L.~Koppens, J.~A.~Folk, J.~M.~Elzerman, R.~Hanson, L.~H.~Willems van Beveren, I.~T.~Vink, H.~P.~Tranitz, W.~Wegscheider, L.~P.~Kouwenhoven, and L.~M.~K.~Vandersypen, Science {\bf 309}, 1346 (2005).
\bibitem{Bayer} A.~Greilich, A.~Shabaev, D.~R.~Yacovlev, Al.~L.~Efros, I.~A.~Yugova, D.~Reuter, A.~D.~Wielck, and M.~Bayer, Science {\bf 317}, 1896 (2007).
\bibitem{Tarucha} Jonathan Baugh, Yosuke Kitamura, Keiji Ono, and Seigo Tarucha, Phys.~Rev.~Lett.~{\bf 99} 096804 (2007).
\bibitem{Steel} Xiadong Xu, Wang Yao, Bo Sun, Duncan G.~Steel, Allan S.~Bracker, Daniel Gammon, and L.~J.~Sham, Nature {\bf 459}, 1105 (2009).
\bibitem{Coish} W.~A.~Coish and Daniel Loss, Phys.~Rev.~B {\bf 72}, 125337 (2005).
\bibitem{Klauser} D.~Klauser, W.~A.~Coish, and Daniel Loss, Phys.~Rev.~B {\bf 73}, 205302 (2006).
\bibitem{Yao} W.~Yao, R.~B.~Liu, and L.~J.~Sham, Phys.~Rev.~Lett.~{\bf 98}, 077602 (2007).
\bibitem{Liu} Wen Yang and Ren-Bao Liu, Phys.~Rev.~B {\bf 78}, 085315 (2008); {\it ibid} {\bf 79}, 115320 (2009).
\bibitem{Witzel} W.~M.~Witzel, and S.~Das~Sarma, Phys.~Rev.~B {\bf 76}, 241303(R) (2007).
\bibitem{Witzel2} L.~Cywinski, W.~M.~Witzel, and S.~Das Sarma, Phys.~Rev.~B {\bf 79}, 245314 (2009).
\bibitem{Deng} C.~Deng and X.~Hu, Phys.~Rev.~B {\bf 72}, 165333 (2005); {\it ibid} {\bf 73}, 241303(R) (2006).
\bibitem{Lukin} Liang Jiang, M.~V.~Gurudev Dutt, Emre Togan, Lily Childress, Paola Cappellaro, Jacob M.~Taylor, and Mikhail D.~Lukin, Phys.~Rev.~Lett.~{\bf 100}, 073001 (2008).
\bibitem{BurLos} G.~Burkard, D.~Loss, and D.~P.~DiVincenzo, Phys.~Rev.~B {\bf 59}, 2070 (1999).
\bibitem{HuSSC} S.~Das Sarma, Rogerio de Sousa, Xuedong Hu, Belita Koiller, Solid State Comm.~{\bf 133}, 737 (2005).
\bibitem{Ramon} Guy Ramon and Xuedong Hu, Phys.~Rev.~B 75, 161301(R) (2007).
\bibitem{PettaPol} J.~R.~Petta, J.~M.~Taylor, A.~C.~Johnson, A.~Yacoby, M.~D.~Lukin, C.~M.~Marcus, M.~P.~Hanson, and A.~C.~Gossard, Phys.~Rev.~Lett.~{\bf 100}, 067601 (2008).
\bibitem{Reilly} D.~J.~Reilly, J.~M.~Taylor, J.~R.~Petta, C.~M.~Marcus, M.~P.~Hanson, and A.~C.~Gossard, Science {\bf 321}, 817 (2008).
\bibitem{Foletti} S.~Foletti, J.~martin, M.~Dolev, D.~Mahalu, V.~Umansky, A.~Yacoby, arXiv:0801.3613 [cond-mat.mes-hall] (unpublished).
\bibitem{Hu} Xuedong Hu and S.~Das Sarma, Phys.~Rev.~Lett.~{\bf 96}, 100501 (2006).
\bibitem{Culcer} Dimitrie Culcer, Xuedong Hu, and S.~Das Sarma, Appl.~Phys.~Lett. {\bf 95}, 073102 (2009).
\bibitem{Pioro} M.~Pioro-Ladri\`{e}re, John H.~Davies, A.~R.~Long, A.~S.~Sachrajda, Louis Gaudreau, P.~Zawadzki, J.~Lapointe, J.~Gupta, Z.~Wasilewski, and S.~Studenikin, Phys.~Rev.~B {\bf 72}, 115331 (2005).
\bibitem{Jung} S.~W.~Jung, T.~Fujisawa, Y.~Hirayama, and Y.~H.~Jeong, Appl.~Phys.~Lett.~{\bf 85}, 768 (2004).
\bibitem{Taubert} D.~Taubert, M.~Pioro-Ladri\`{e}re, D.~Schr\"{o}er, D.~Harbusch, A.~S.~Sachrajda, and S.~Ludwig, Phys.~Rev.~Lett.~{\bf 100}, 176805 (2008).
\bibitem{Taylor} J.~M.~Taylor, H.~-A.~Engel, W. D\"{u}r, A.~Yacoby, C.~M.~Marcus, P.~Zoller and M.~D.~Lukin, Nature Physics {\bf 1}, 177 (2005).
\bibitem{Puerto} Irene Puerto Gimenez, Chang-Yu Hsieh, Marek Korkusinski, and Pawel Hawrylak, Phys.~Rev.~B {\bf 79}, 205311 (2009).
\bibitem{Roszak} K.~Roszak and P.~Machnikowski, arXiv:0903.0783 [cond-mat.mes-hall] (unpublished).
\bibitem{HuArx} Xuedong Hu, cond-mat arXiv:0903.1080 [cond-mat.mes-hall] (unpublished).
\bibitem{Astafiev} O.~Astafiev, Yu.~A.~Pashkin, Y.~Nakamura, T.~Yamamoto, and J.~S.~Tsai, Phys.~Rev.~Lett.~{\bf 93}, 267007 (2004); {\it ibid} {\bf 96}, 137001 (2006).
\bibitem{Martinis} John M.~Martinis, K.~B.~Cooper, R.~McDermott, Matthias Steffen, Markus Ansmann, K.~D.~Osborn, K.~Cicak, Seongshik Oh, D.~P.~Pappas, R.~W.~Simmonds, and Clare C.~Yu, Phys.~Rev.~Lett.~{\bf 95}, 210503 (2005).
\bibitem{Shnirman} Alexander Shnirman, Gerd Sch\"{o}n, Ivar Martin, and Yuriy Makhlin, Phys.~Rev.~Lett.~{\bf 94}, 127002 (2005).
\bibitem{Paladino} E.~Paladino, M.~Sassetti, G.~Falci, and U.~Weiss, Phys.~Rev.~B {\bf 77}, 041303(R) (2008).
\bibitem{Bergli} J.~Bergli, Y.~M.~Galperin, and B.~L.~Altshuler, New Journal of Physics {\bf 11}, 025002 (2009).
\bibitem{Paladino1} E.~Paladino, L.~Faoro, G.~Falci, and Rosario Fazio, Phys.~Rev.~Lett.~{\bf 88}, 228304 (2002).
\bibitem{Faoro} Lara Faoro, Joakim Bergli, Boris L.~Altshuler, and Y.~M.~Galperin, Phys.~Rev.~Lett.~{\bf 95}, 046805 (2005).
\bibitem{delta} We excluded a constant term that does not play any role in the system dynamics.
\bibitem{HilWoo} Scott Hill and William K.~Wootters, Phys.~Rev.~Lett.~{\bf 78}, 5022 (1997).
\bibitem{bias} In refering to the double dot bias we adopt the convention of zero bias at the anticrossing point.
\bibitem{spherical} We do not expect that a choice of a spheical symmetric TLS wavefunction will induce any qualitative change in our results.
\bibitem{TLStun} When tunneling between the TLS centers is introduced the TLS eigenstates become $(L \pm R)/\sqrt{2}$ (for $\omega_T=0$) and contributions from off-diagonal TLS states become appreciable. The qubit-TLS coupling then produces additional terms in the interaction Hamiltonian (Eq.~\ref{Hint}) of the form $\sigma_x^T \otimes \sigma_z^Q$, $\sigma_x^T \otimes I^Q$. Assuming the TLS is unbiased there is no TLS dipole term and the leading contribution comes from the TLS quadrupole term, which is calculated to be very small for the TLS tunneling values assumed in this paper.
\bibitem{Jackson} See, e.g., chapter 4 in J.~D.~Jackson, {\em Classical Electrodynamics, 2nd edition} (John Wiley, New York, 1975).
\bibitem{monopole} Notice that while the qubit holds two electrons, its charge density corresponds to single-particle operators. The two-particle qubit states are considered at a later stage, when evaluating the Coulomb matrix elements, Eqs.~(\ref{ST}).
\bibitem{PettaPRL} J.~R.~Petta, A.~C.~Johnson, C.~M.~Marcus, M.~P.~Hanson, A.~C.~Gossard, Phys.~Rev.~Lett.~{\bf 93}, 186802 (2004).
\bibitem{HanBur} R.~Hanson and G.~Burkard, Phys.~Rev.~Lett.~{\bf 98}, 050502 (2007).
\bibitem{Switch} We have verified that switching times $\lesssim 1$ ns, comparable to experimentally used bias sweep times,\cite{Petta} do not contribute appreciably to gate errors in most cases, and in any case can be corrected by a careful pulse design. To isolate the effects of qubit-TLS coupling we have nevertheless considered very short switching times ($\sim 1$ ps).
\bibitem{Overdamp} Throughout most of the parameter ranges presented in Figure \ref{Fig7n}, $t_T>\Gamma$ thus the TLS is not overdamped exclugin a direct coupling of its reservoir to the qubit.
\bibitem{beta0} Higher order contributions to $\beta$ in the multipole expansion may have an opposite sign to that of the leading term. Typically, their magnitude will be comparable at shorter distances ($R\lesssim 25$ nm), where the expansion converges slowly and yet higher order contributions are needed for an accurate estimate of the coupling.
\bibitem{para} The geometry depicted in Fig.~\ref{Fig8n}b yields $\beta/\gamma \sim 1$ and thus yields  considerably larger gate errors and dephsing as compared with the parallel-axis geometry.
\bibitem{F4} Notice that the function $F_4$ in ref.~\onlinecite{BurLos} can be simplified as $F_4=c\sqrt{b}e^{d^2/4b} I_0 (d^2/4b)$, where $c=\sqrt{\pi/2}(e^2/\kappa a_B)/\hbar \omega_0$ is the ratio between Coulomb and confining energies, and $I_0$ is the zeroth order modified Bessel function of the first kind.
\end{thebibliography}

\end{document}